\documentclass{ws-jai}
\usepackage[flushleft]{threeparttable}

\begin{document}

\catchline{}{}{}{}{}

\markboth{S. B. Pandey et al.}{4K$\times$4K CCD Imager for the 3.6m DOT: Recent up-gradations and results}

\title{4K$\times$4K CCD Imager for the 3.6m DOT: Recent up-gradations and results}

\author{S. B. Pandey$^{1*}$, Amit Kumar$^{1,2**}$,  B. K. Reddy$^{1}$, S. Yadav$^{1}$, N. Nanjappa$^{1}$, Amar Aryan$^{1,3}$, Rahul Gupta$^{1,3}$, Neelam Panwar$^{1}$, and R. K. S. Yadav$^{1}$}

\address{
$^{1}$Aryabhatta Research Institute of Observational Sciences (ARIES), Manora Peak, Nainital, Uttarakhand, India, 263001\\
$^{2}$School of Studies in Physics and Astrophysics, Pandit Ravishankar Shukla University, Raipur 492010, Chhattisgarh, India\\
$^{3}$Department of Physics, Deen Dayal Upadhyaya Gorakhpur University, Gorakhpur 273009, Uttar Pradesh, India\\
}
\maketitle

\corres{$^{*}$S. B. Pandey, shashi@aries.res.in; $^{**}$A. Kumar, amitkundu515@gmail.com}

\begin{history}
\received{(2022)};
\revised{(xxxx)};
\accepted{(xxxx)};
\end{history}

\begin{abstract}
The 4K$\times$4K CCD Imager is the first light instrument for the 3.6m Devasthal Optical Telescope and is producing broad-band imaging observations of many Galactic and extra-galactic sources since 2015-2016. Capabilities of the CCD Imager are demonstrated recently through several publications using the well-calibrated multi-band deep photometric results as expected from other similar facilities globally. In this article, we summarize some of the recent up-gradations made to improve the Imager, i.e., mounting the new filter wheel casing, replacing stray light baffles and discussing the fringe pattern corrections in redder filters. Some of the new science initiatives like galaxy-embedded faint point sources including WR stars and the observations of low surface brightness galaxy clusters are also discussed.
\end{abstract}

\keywords{Instrumentation: CCD photometry; Methods: Optical observations; Data analysis.}

\section{Introduction}

\noindent Location of the modern and active optics based 3.6m Devasthal Optical Telescope (DOT) is well-suited for the time-critical optical-near infrared (NIR) observations \citep{Pandey2016, Brijesh2018}. Devasthal, the new observing station of Aryabhatta Research Institute of Observational Sciences (ARIES), Nainital at an altitude of $\sim$ 2450m (longitude 79.7E, latitude 29.4N), \citep{Sagar2000}, having advantages like dark skies, sub-arcsec seeing conditions has now been established as a world-class astronomical site hosting other observational facilities including 1.3m Devasthal Fast Optical Telescope, and recently commissioned 4m International Liquid Mirror Telescope too. Among suit of many other back-end instruments, 4K$\times$4K CCD Imager is the first light axial-port imaging instrument for the 3.6m DOT and was made functional in late 2015 by observing the very first image as a crab nebula observed on 11$^{th}$ December 2015 as reported by \citet{Pandey2016}.  The 4K$\times$4K CCD Imager is designed and developed in-house to be mounted at the main/axial port of the 3.6m telescope as one of the first-light instruments. The beam of the telescope is f/9, used without any focal reducer and has a plate-scale of 6.4$''$/mm. Details about the CCD characterization, photometric calibration along with the site characterization (the calculation of extinction coefficients and sky brightness) using the instrument mounted at the axial port of the 3.6m DOT have been studied by \citet{Pandey2018a, Kumar2022a}. In recent times, the 4K$\times$4K CCD Imager has been utilized as a general multi-band deep imaging instrument to study various kinds of Galactic and extra-galactic astronomical sources such as star clusters \citep{Lata2019, Panwar2022, 2022JApA...43...31S}, cosmic energetic transients \citep{Pandey2019, Dastidar2019, Kumar2020GCN, Kumar2021_ank, Aryan2021, 2021RMxAC..53..215A, Gupta2021, 2021arXiv211111795G, 2021GCN.31299....1G, Pandey2021, 2022MNRAS.511.1694G, 2022arXiv220513940G, 2022MNRAS.513.2777K, 2022JApA...43...87A, 2022MNRAS.517.1750A, 2022arXiv220600950K}, active galactic nucleus \citep{Ojha2022}, and many others.

In this article, we present some recent updates about the attempts made towards upgrading the existing CCD Imager to improve upon some of the teething problems encountered over several observing seasons. Some of the important ones are discussed in this study that includes improvements made towards old baffle design having problems of partial vignetting, up-gradations made towards motorized filter wheels to improve the filter repositioning and attempts to remove possible fringing issues for broad-band near-infrared filters. The results reported in this article are based on experiences over several observing seasons. After incorporating the specified improvements, the results obtained from the first light instrument have been improved considerably and have also helped to seek lessons for other back-end instruments for the 3.6m DOT. We have organized this article as follows: In section \ref{Baffle}, we present the details of the vignetting and ghost along with the baffle design of the 4K$\times$4K CCD Imager, followed by the description of mechanical design and analysis in section \ref{machanical}. In section \ref{filters}, we discuss the up-gradations of the motorized filter wheel. In section \ref{Fringr}, we present the method for the fringe correction of red-end filters. In section \ref{science}, we mention some of the recent science cases studied using 4K$\times$4K CCD Imager and finally, a summary of the paper is given in section \ref{summary}.

\section{Vignetting/Ghost analysis of the 4K$\times$4K CCD Imager}
\label{Baffle}

\begin{figure*}[!ht]
\centering
\includegraphics[scale=0.155]{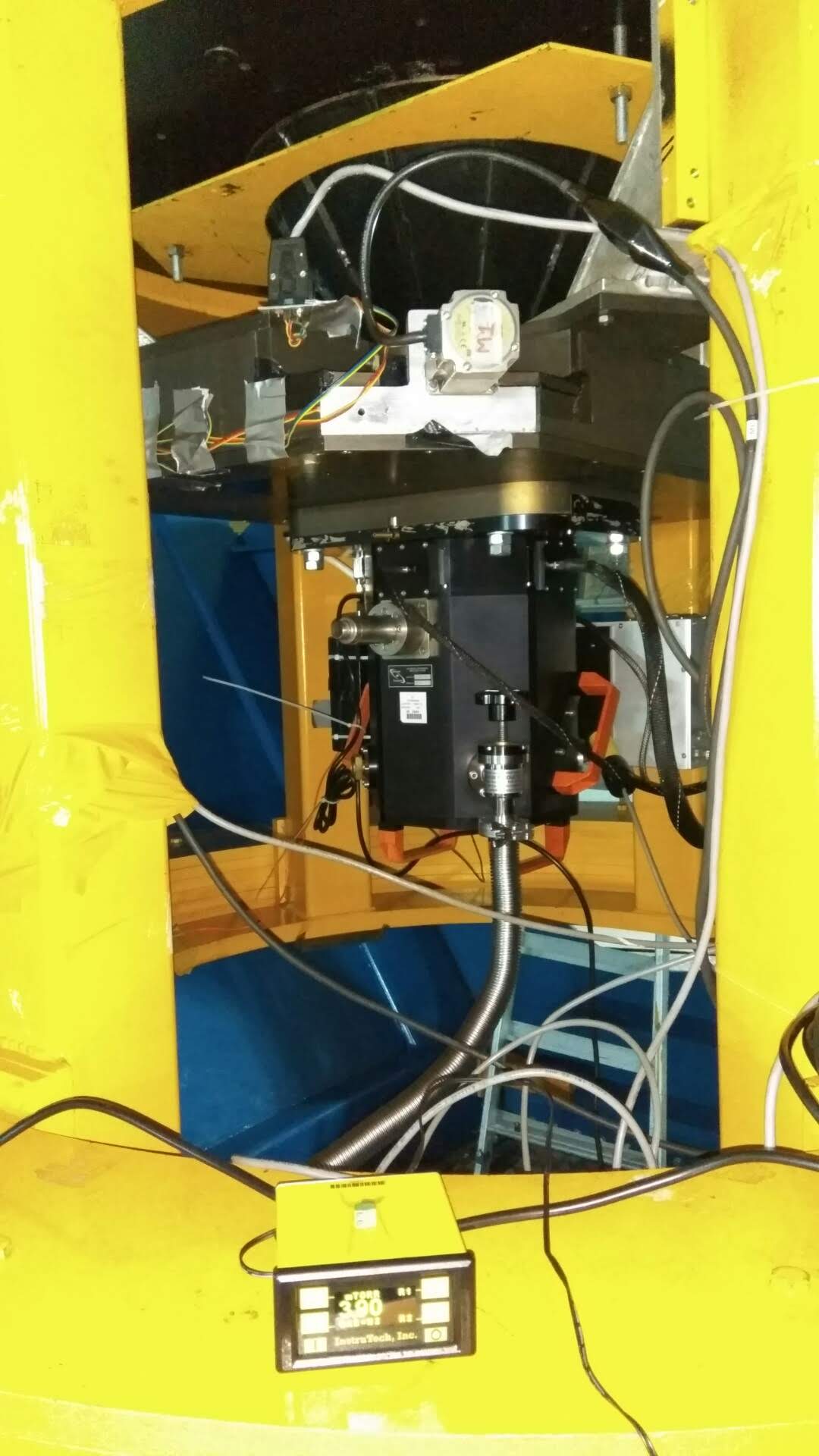}
\includegraphics[scale=0.38]{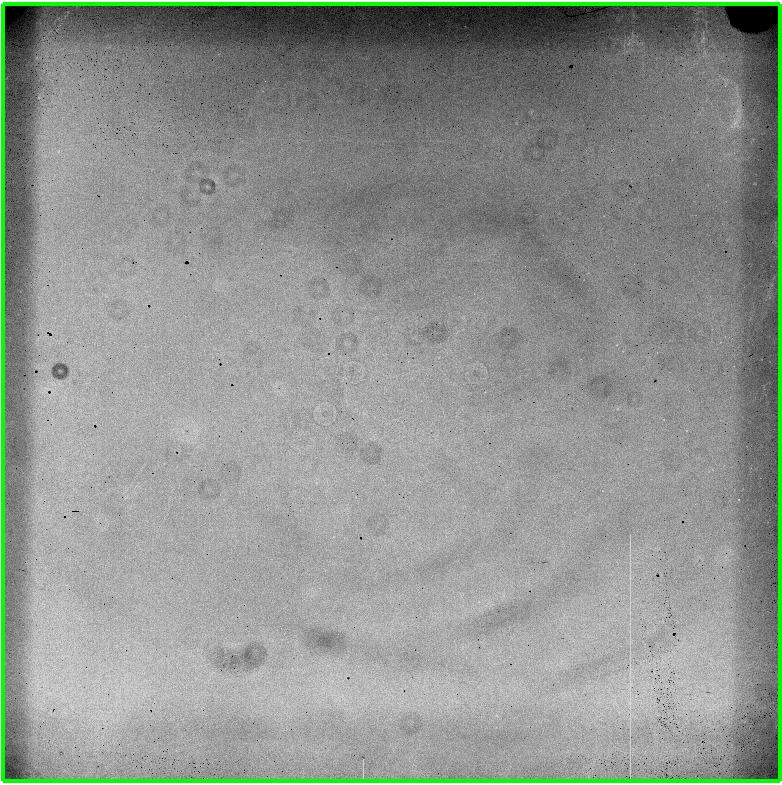}
\caption{Left panel: 4K$\times$4K CCD Imager mounted on the axial port of 3.6m DOT. Right panel: vignetting feature at the four sides of CCD Imager.}\label{FIGONE}
\end{figure*}

The 4K$\times$4K CCD Imager was mounted on the axial port of 3.6m DOT (shown in the left panel of Figure~\ref{FIGONE}). It has a fully assembled Imager set-up along with automated two filter wheels having Bessel $UBVRI$ and SDSS $ugriz$ filters. Each filter is assembled in its respective positions with the help of a Teflon cover. The actual size of each filter is 90$\times$90mm, and the clear aperture is 85$\times$85mm; the remaining 5mm is used for mounting with a Teflon cover as a collar to hold the filter. Shutter was integrated with the CCD flange, and this assembly was mounted with the Imager. The total assembly was mounted to the dummy of the telescope with the help of three arm structures. There has been a gap of around 300mm between the telescope flange and the Imager (which means filter wheel entrance). Initially, a conical baffle was kept between the telescope flange and the Imager by matching the hole size. Images were taken using this configuration. It was observed that there was a shadow area at the four corners of the CCD, as shown in the right panel of Figure \ref{FIGONE}.

Images were taken in various combinations using Imager to investigate the shadow/ghost feature. The width of the shadow region was less when there were no filters, and increased in the Filter-Clear combination. Dark frames were taken by closing the dome slit and switching off the lights inside the dome. The same feature was observed with less shadow width. This feature is visible in the on sky and flat images also. The same feature was observed even when the entire imager assembly shifted downwards by 7.5mm. The feature was visible in almost all images.

\subsection{Zemax analysis to verify the vignetting and ghost factors}

\begin{figure*}[ht!]
\centering
\includegraphics[scale=0.35]{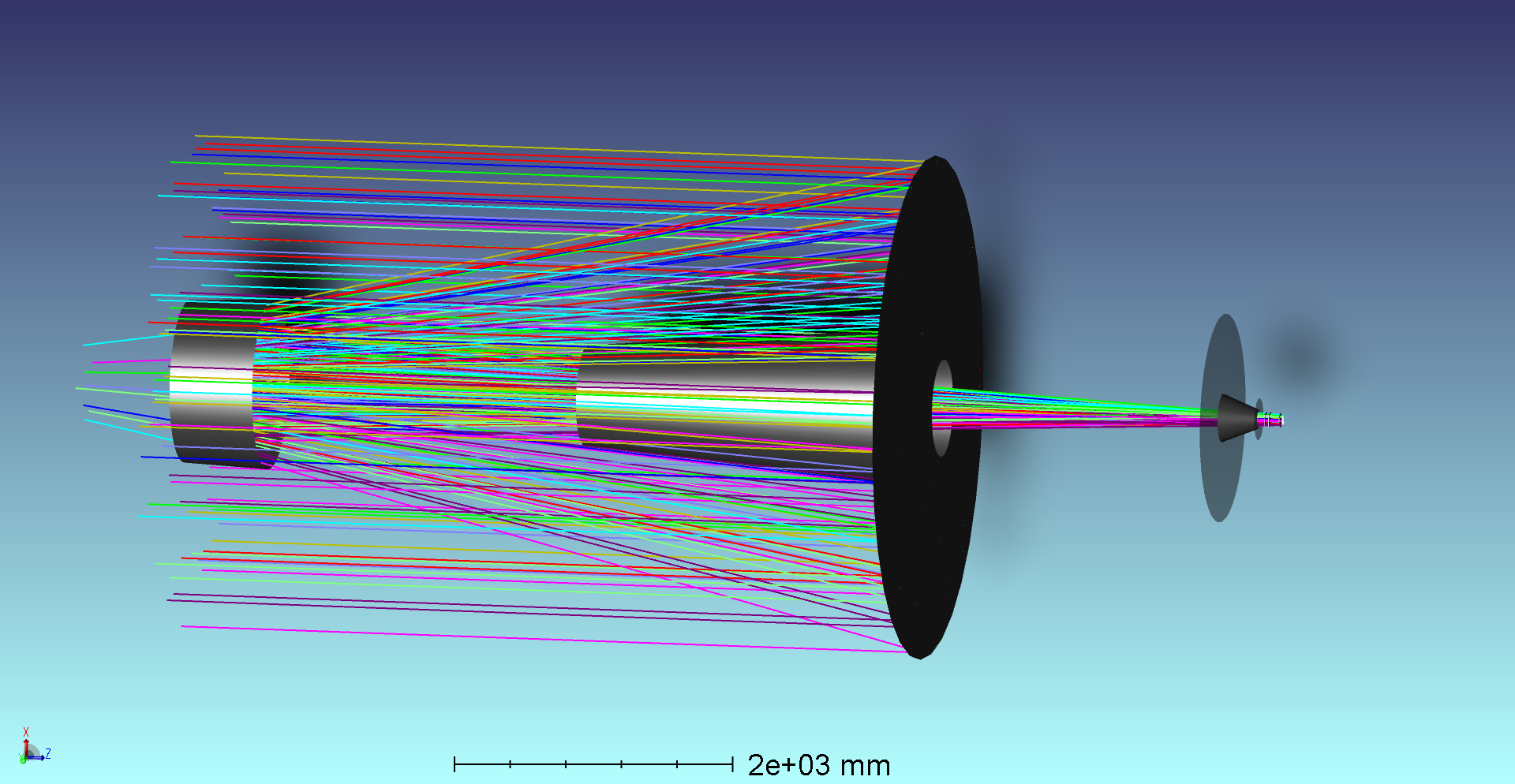}
\includegraphics[scale=0.35]{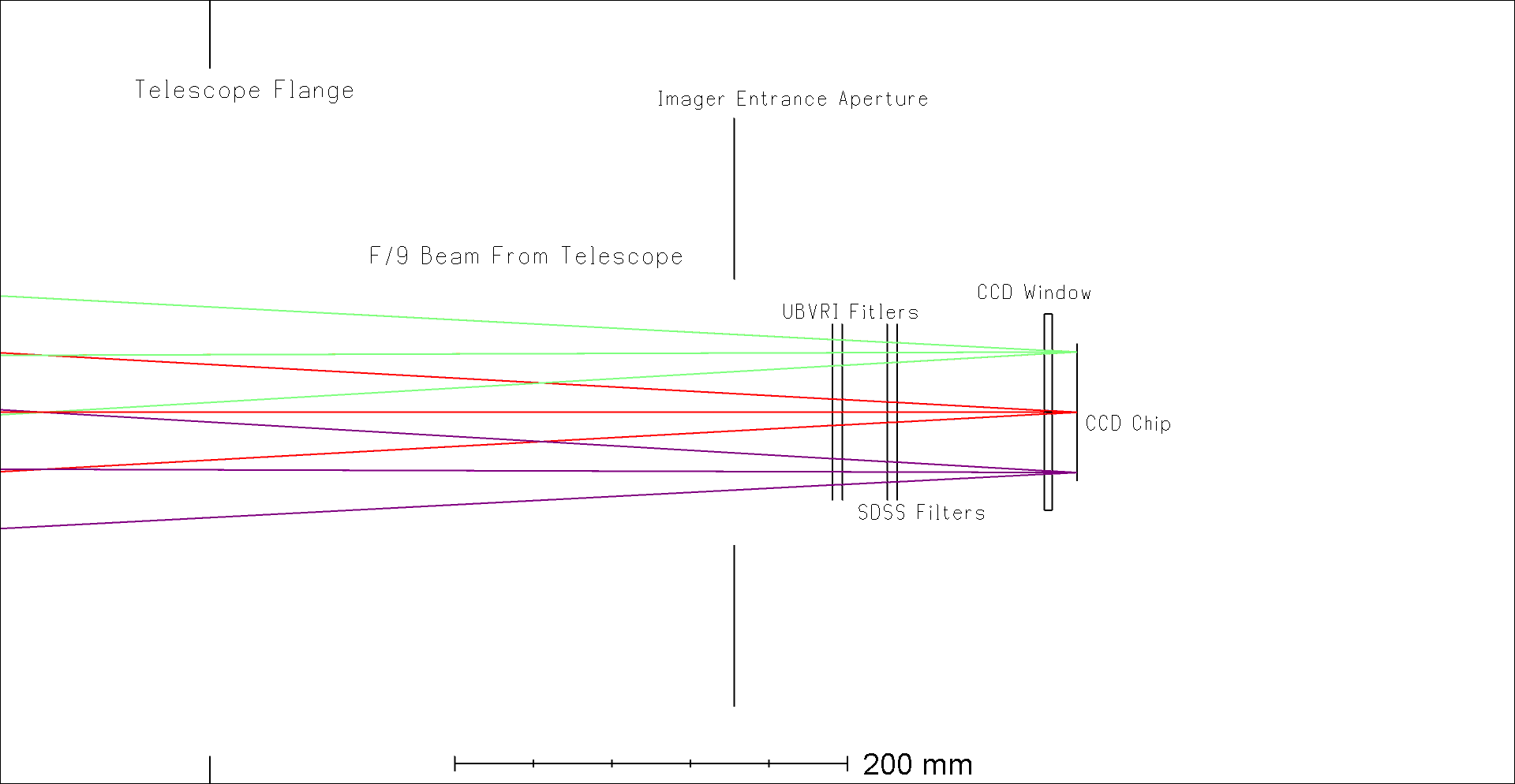}
\caption{Top panel: the shaded model of 3.6m telescope with Imager optics. Bottom panel: optical layout of the 4K$\times$4K CCD Imager system.}\label{FIGTWO}
\end{figure*}

To analyze possible issues like the vignetting and ghost images, the ZEMAX file is made with built specifications of the telescope and image plane kept at its best focus for the field of 6$^\prime$.52$\times$6$^\prime$.52 to cover full CCD chip size of 61.44$\times$61.44mm. The shaded model of the 3.6m telescope with Imager optics is shown in the top panel of Figure \ref{FIGTWO}. The filter wheel and CCD window are kept in the optical path of the telescope (see the bottom panel of Figure \ref{FIGTWO} for the optical layout of the Imager). The ghost effect on the CCD chip is analyzed due to the filter and CCD window. Filter (Fused silica) with 5mm thickness and CCD window (Fused silica) of thickness 4mm are considered for this analysis. The space between the filter front surface and the CCD image plane is 123mm, and between the CCD window front surface and the image plane is 15.3mm.\\

\begin{figure*}[ht!]
\centering
\includegraphics[scale=0.25]{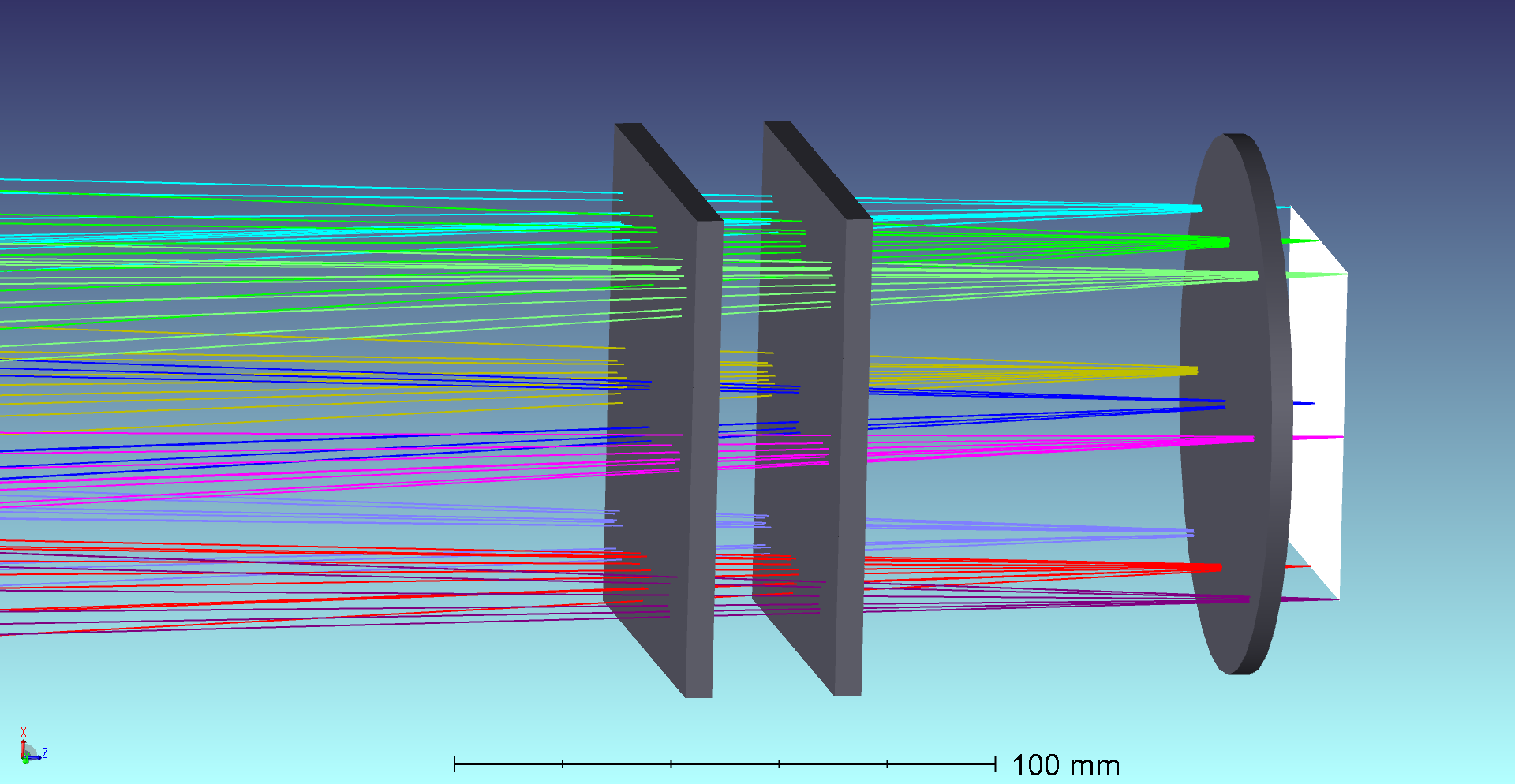}
\includegraphics[scale=0.4]{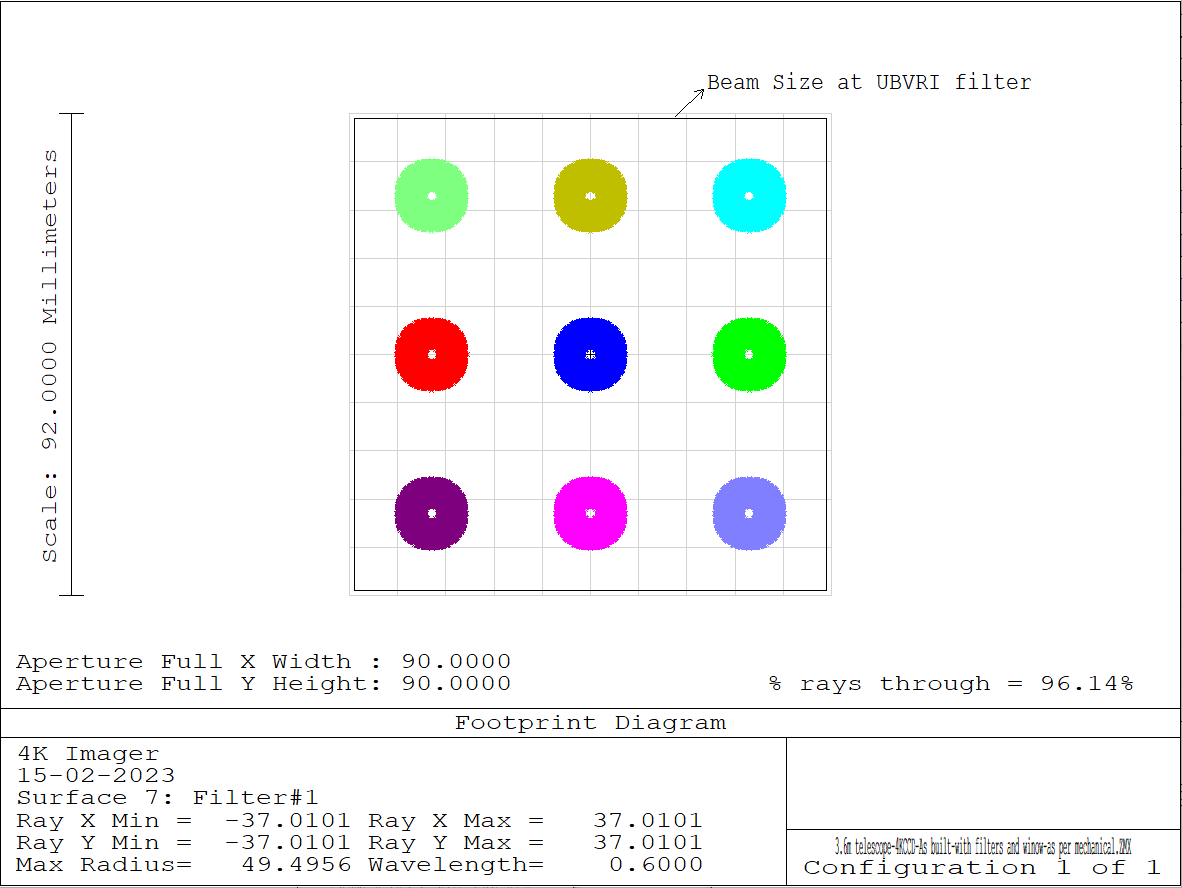}
\caption{Top panel: 3D layout of beam path of the 4K$\times$4K CCD Imager system. Bottom panel: full beam size on filter surface as seen by the ZEMAX file for the set-up.}\label{FIGTHREE}
\end{figure*}

The vignetting is verified in ZEMAX sequential mode as per original dimensions. The maximum beam size on the filter with theoretical distances is around 75$\times$75mm. The 3D layout of the Imager's beam path and optical layout of the full beam size on the filter surface is shown in the top and bottom panels of Figure \ref{FIGTHREE}, respectively. No vignetting due to filter size is observed for the given field of view by assuming that there is no significant mismatch in mechanical distances.\\

\begin{figure*}[!t]
\centering
\includegraphics[scale=0.55]{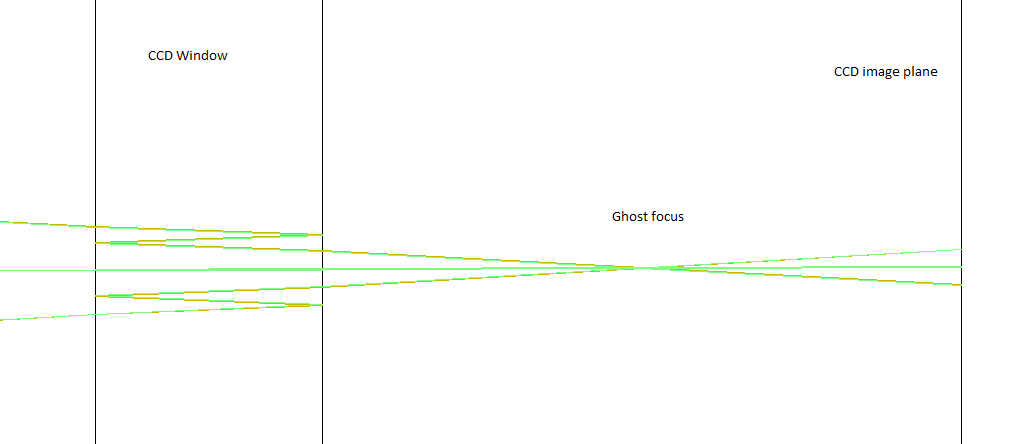}
\includegraphics[scale=0.45]{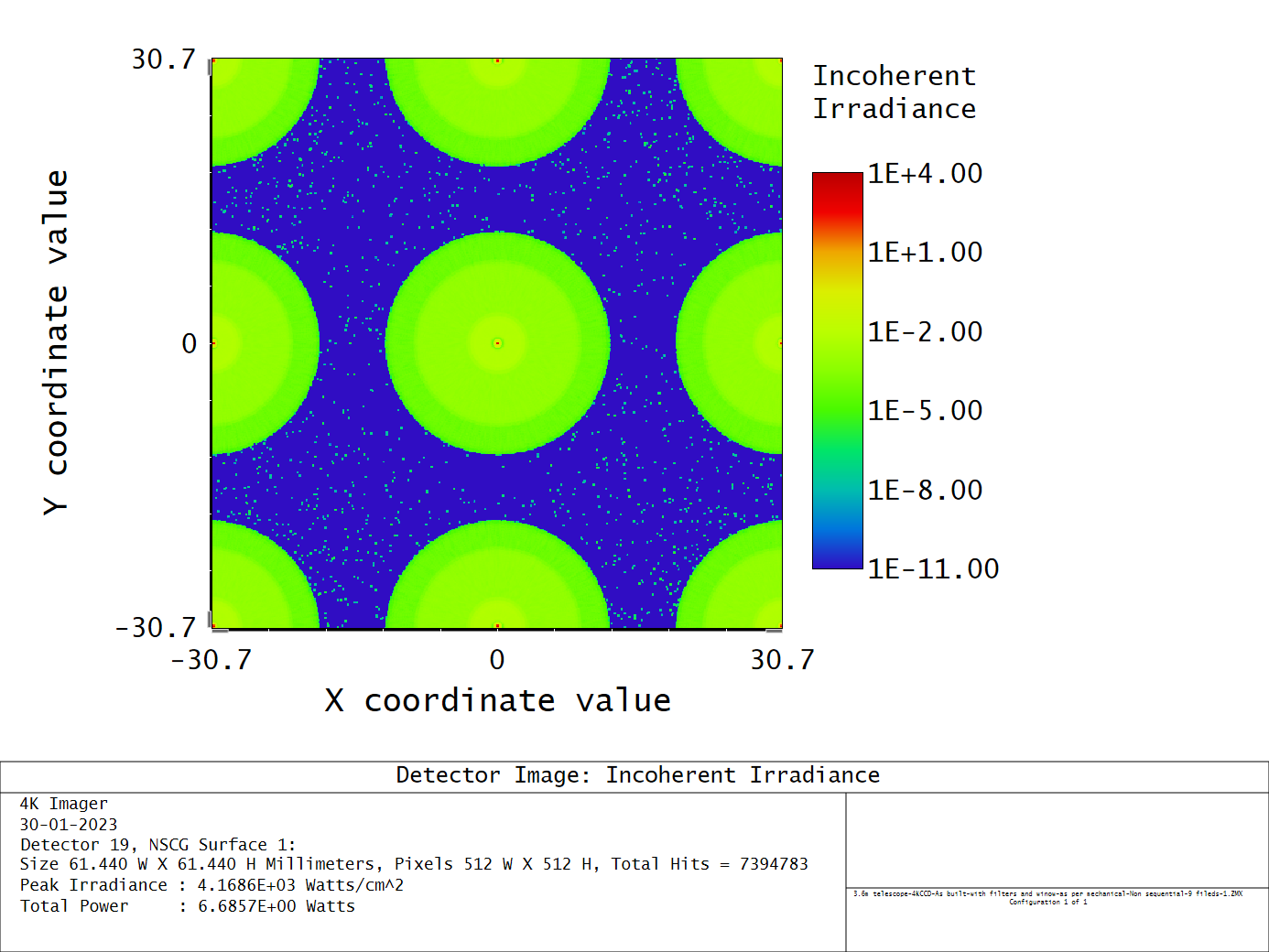}
\caption{Top panel: ghost image due to internal reflections of CCD window in sequential mode. Bottom Panel: ghost image due to filters, CCD window in non-sequential mode.}\label{FIGFOUR}
\end{figure*}

The ghost analysis is carried out to see the unwanted ghost shadows coming from the internal optical components. The ghost images can be generated by two major groups of reflections. One is due to the reflections inside the optical elements like the filter and CCD window, including the CCD chip plane. Another is due to the reflections between CCD window/imaged planes and filter elements. A schematic diagram presenting internal reflections of the CCD window as a main cause of the ghost image is shown in the top panel of Figure~\ref{FIGFOUR}. The ZEMAX routine "ghost focus generator" was used for this analysis, and the given results were analyzed. Double bounce (2 reflections) ghosts are generated up to the CCD chip considering the telescope's primary mirror, secondary mirror, filter, and CCD window. The final results are analyzed in terms of the most critical ghost-focalized point (closest ghost focus). After the analysis, it is observed that the most critical ghost is due to the CCD window and filter. The ghost focalized point due to the CCD window is around 5.14mm before from the CCD image plane. The ghost focalized point due to the filter is around 6.51mm before from the CCD image plane. 99\% anti-reflection coating was considered for the filter, CCD window, and CCD chip. So the ghost image diameter due to the window on the CCD chip is around 570 microns which is 360 times greater than the characteristic size of the image spot (size of two pixels). It means that the ratio between the lightning is 360/.0001, which is 3.6$\times$10$^{6}$. The same was simulated in the non-sequential mode of ZEMAX (as shown in the bottom panel of Figure~\ref{FIGFOUR}). The detector viewer figure shows the intensity values (in log-15 scale) of actual star images (red coloured) and their corresponding ghost pupils (green clouded). These ghost images due to the CCD window or filter are insignificant, and these are not visible with the original image in regular observations by considering the dynamic range of CCD and magnitude of objects.

\begin{figure*}[ht!]
\centering
\includegraphics[scale=0.25]{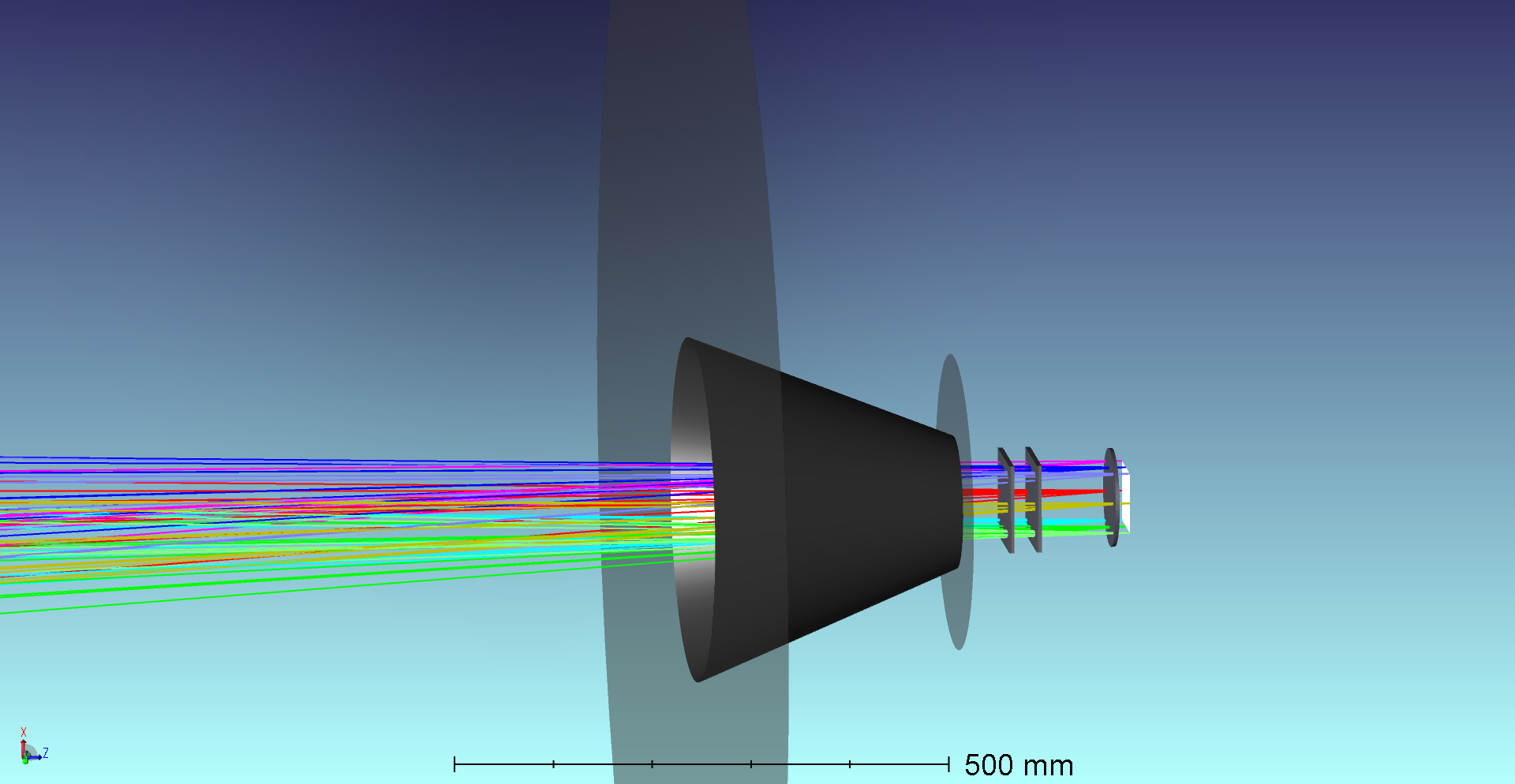}
\includegraphics[scale=0.3]{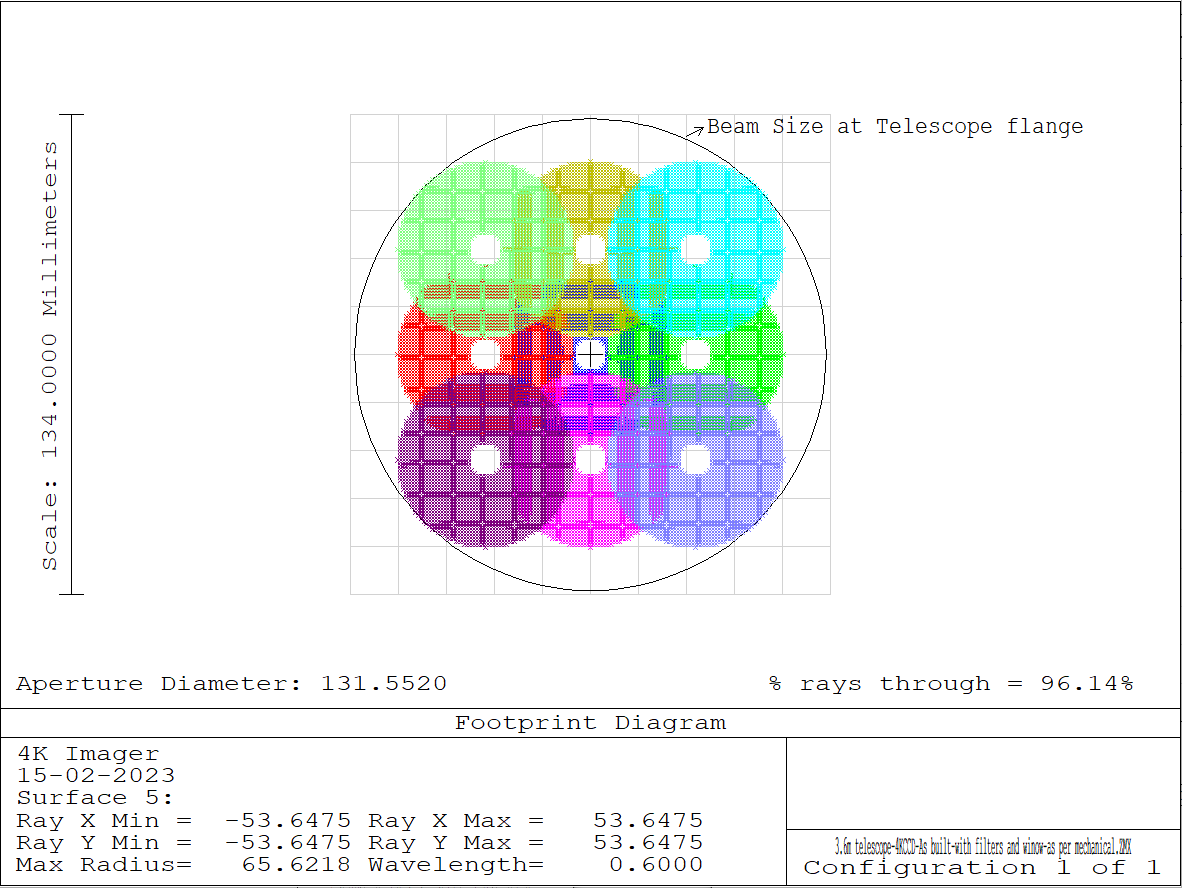}
\includegraphics[scale=0.3]{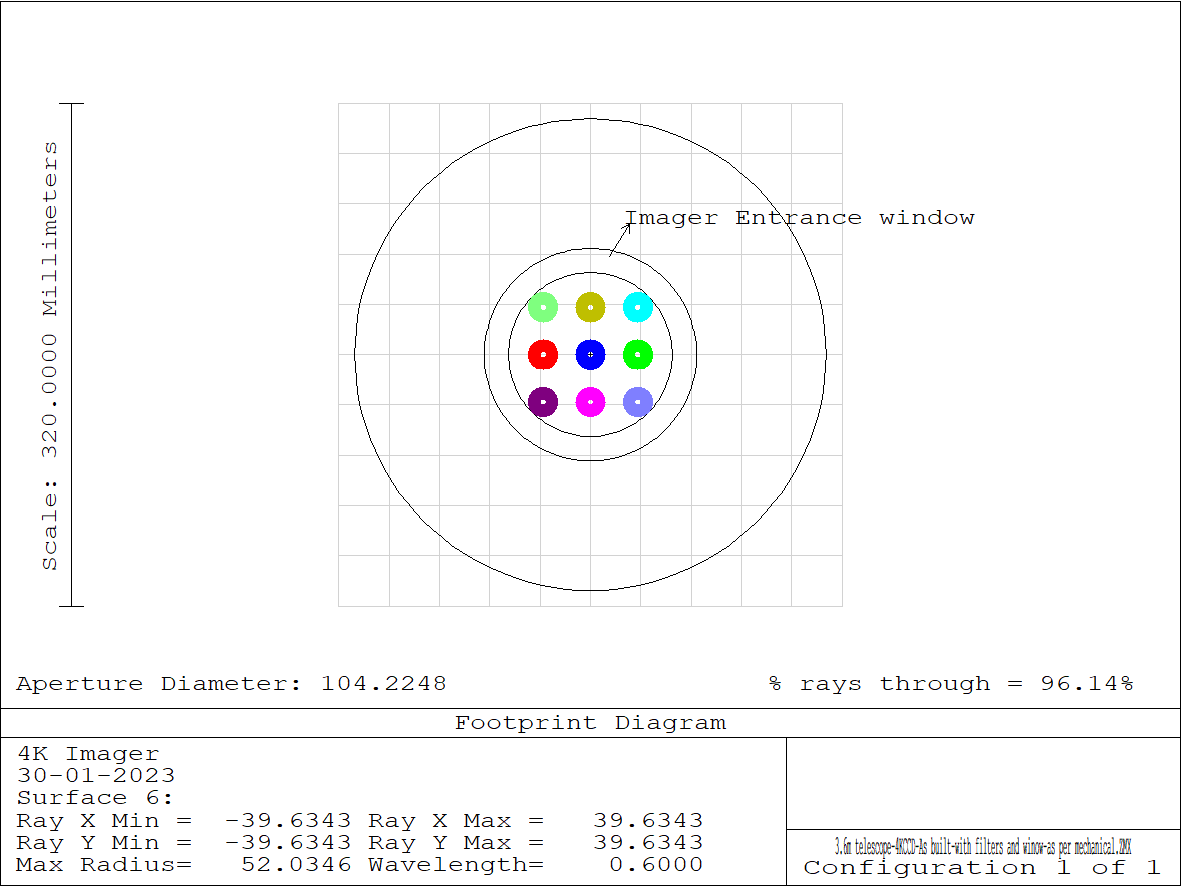}
\caption{Top panel: existing baffle view. Middle panel: beam size at telescope flange. Bottom panel: beam size at Imager entrance point.}\label{FIGFIVE}
\end{figure*}

\begin{figure*}[!t]
\centering
\includegraphics[scale=0.28]{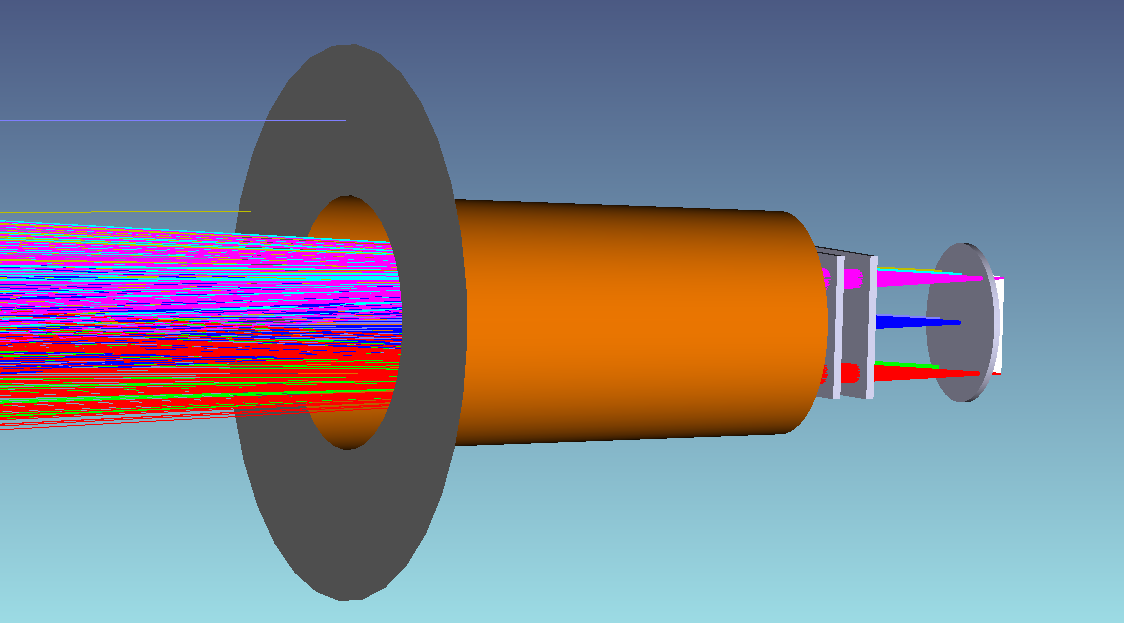}
\includegraphics[scale=1.0]{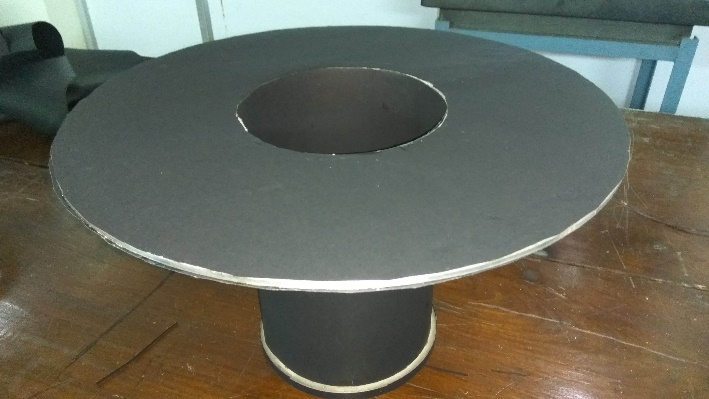}
\caption{Left panel: close view of baffle simulated in ZEMAX non-sequential mode. Right panel: new baffle as finally mounted with the CCD Imager.}\label{FIGEIGHT}
\end{figure*}

It was concluded from the ZEMAX simulations that the shadow region at four sides of the CCD was neither due to ghost reflections of filters and CCD window nor due to vignetting from the slit wheel slot sizes. It was clear that the issue was not due to the instrument itself; rather, it was from the outside of the instrument setup, especially from the Adapter Rotator Instrument Support Structure (ARISS) telescope flange. The existing baffle was investigated (see the top panel of Figure~\ref{FIGFIVE}), and it was found that the size of the existing baffle was much bigger than the beam size at the telescope flange, as shown in the middle panel of Figure~\ref{FIGFIVE}. Actual beam sizes at the telescope flange and imager entrance aperture are shown in the bottom panel of Figure~\ref{FIGFIVE}. Unwanted light from out of the field was coming through this baffle. To avoid this unwanted light, a new baffle was designed as per the beam size at the telescope flange and imager entrance window by simulating the beam in ZEMAX for a full field of 6.5’$\times$6.5’ (see the right panel of Figure~\ref{FIGEIGHT}). The size of the baffle has been kept a little more than the actual footprint beam sizes by considering the existing Imager setup and its tolerances (an image of the new baffle is shown in the left panel of Figure \ref{FIGEIGHT}).

After concluding the sizes from the simulations, a baffle was made using an existing ms-sheet with the help of the mechanical workshop. A dull black paper was glued to the entire surface of the baffle to restrict the unwanted stray light scattering. This new baffle was replaced with the old baffle of 4K$\times$4K CCD Imager. Later, the 4K$\times$4K CCD Imager was mounted with this new baffle, and images were taken to test the shadow feature at the four sides of the CCD. After integrating this new baffle with the 4K$\times$4K CCD Imager, the shadow feature completely disappeared. \\

\begin{figure*}[!t]
\centering
\includegraphics[scale=0.5]{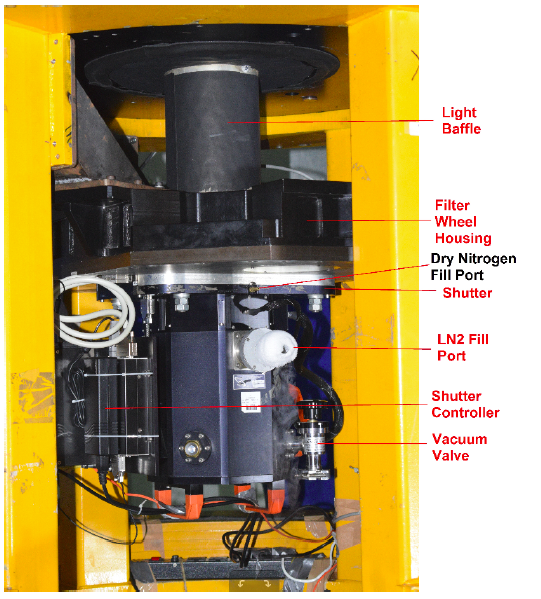}
\includegraphics[scale=0.29]{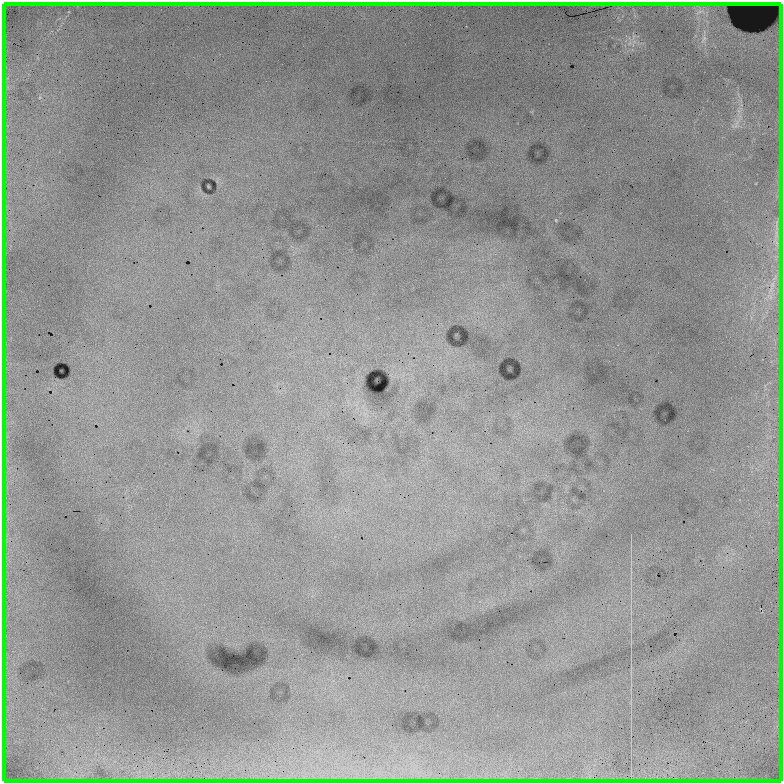}
\caption{Left panel: Imager mounted at the axial port of the 3.6m DOT with the new light baffle. Various parts of the CCD Imager are also marked for clarity; the figure is adapted from \citet{Kumar2022a}. Right panel: vignetting-free flat frame taken after placing new light baffle.}\label{FIGNINE}
\end{figure*}

In all, there is no vignetting feature due to the filter/wheel if we assume all mechanical distances (telescope flange to filters to CCD chip plane) to be accurate. Ghost analysis was carried out to check the ghost effect due to the filters and CCD window. Ghost due to the filters and CCD window being negligible and not visible in the original image. Finally, it was concluded that the shadow region was due to unwanted light from the big opening size of the previous baffle. The problem was resolved after mounting the new baffle designed, manufactured, assembled, and tested within the ARIES workshop. The left and right panels of Figure~\ref{FIGNINE} show the setup of Imager with a new baffle and shadow-free image, respectively.

\section{Mechanical design and analysis of the filter-housing of the 4K$\times$4K Imager}
\label{machanical}

The primary mirror of the 3.6m DOT is actively supported with 69 actuators which correct the surface deformations with wavefront camera feedbacks mounted in the ARISS unit of the telescope. These actuators also provide suitable space at the axial port of the telescope. Later, the 4K$\times$4K CCD Imager was mounted in the year 2015-2016 at Devasthal astronomical site, India. Finally, it was successfully operated to study deep optical photometric observations of galactic and extra-galactic sources \citep{Pandey2018a, Kumar2022a}.

\subsection{Description of Mechanical Design}

\begin{figure*}[!t]
\centering
\includegraphics[scale=0.25]{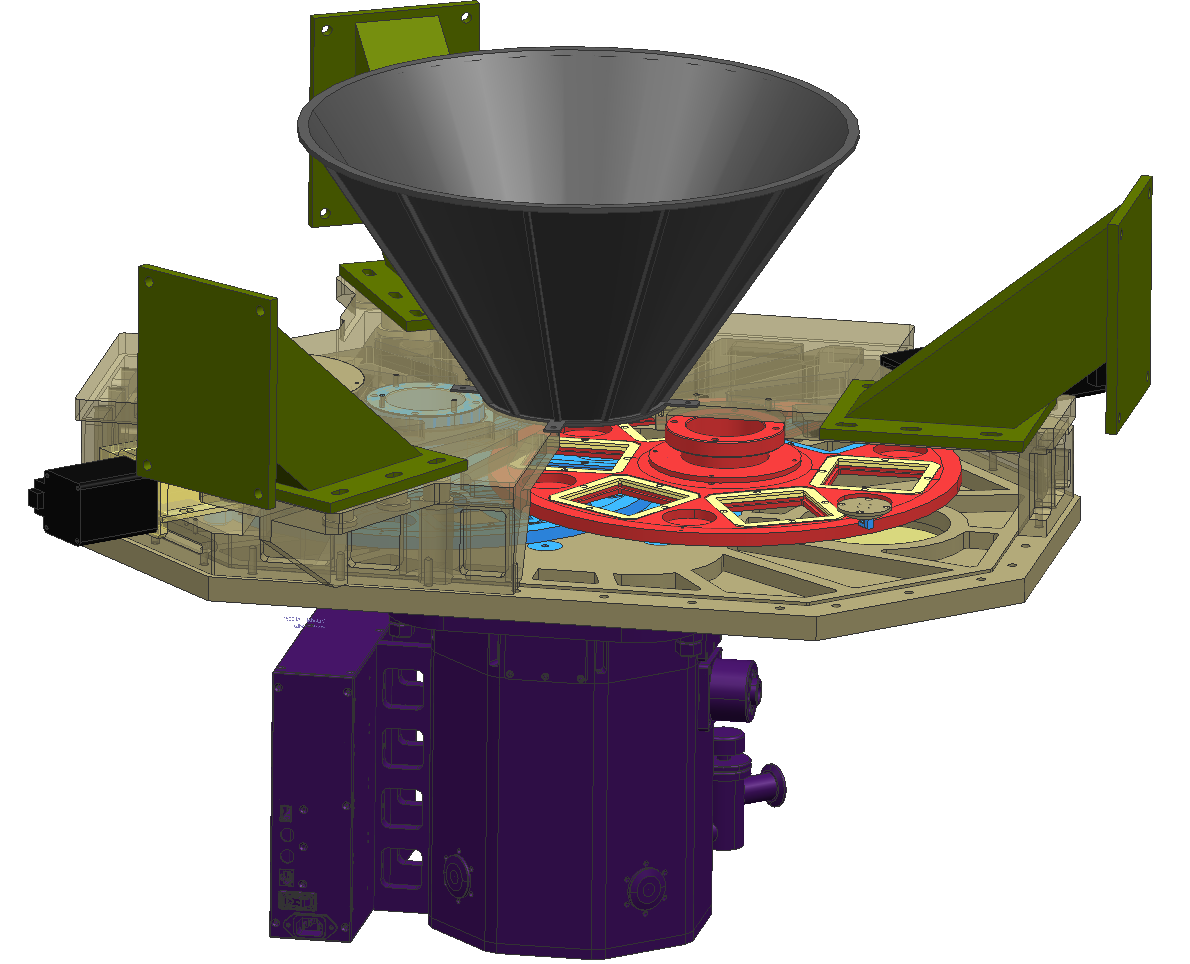}
\includegraphics[scale=0.6]{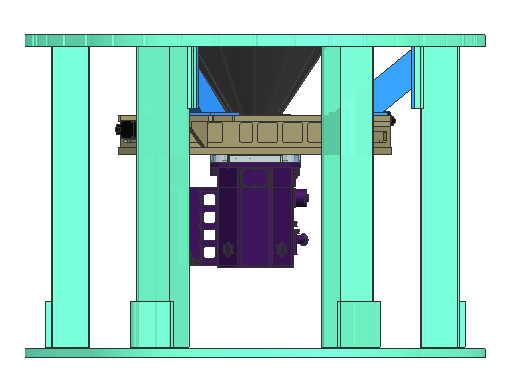}
\caption{Left panel: image representing the Imager housing structure, support arm, filter wheel assembly, and 4K$\times$4K CCD. Right panel: image showing the light baffle + filter assembly + Imager mounted at the axial port of the 3.6m DOT.}\label{FIGTEN}
\end{figure*}

The Imager has a housing structure, support arms, filter wheel assembly, and 4K$\times$4K CCD. The housing is the main structure of the Imager instrument, which houses the two-filter wheels assembly and provides stable support to the CCD camera through a precisely machined bottom plate (see the right and left panels of Figure~\ref{FIGTEN}). The housing structure is a ribbed aluminium cast body of size 850$\times$560$\times$140mm and the bottom plate is an AL6061t6 ribbed machined plate in CNC.

The filter wheel assembly has two filter wheels of diameter 482mm each with drive mechanisms like bearing housings, motors, gearbox and electronics sensor, etc. Each filter wheel has 6 pockets to mount 5 filters of size 90mm$\times$90mm. Both the filter wheels have a $clear$ pocket too. These filter wheels are mounted parallel to each other with a gap of 10mm. The filter wheels are fabricated in Al 6061t6 alloy to reduce the inertia and black anodized to avoid rusting. 
	
The three arms in the Imager instrument provide stable fixation support in the 3.6m DOT dummy structure. The arms are kept at 120 degrees in the dummy structure to simplify the housing structure and provide access for filter replacement without disassembling the instrument. The baffle cover is designed in a later stage to avoid the stray light in the telescope dome, and the conical shape has been changed to a cylindrical shape to avoid light vignetting during observations.

To achieve the quality image in this instrument, the positions of filters in filter wheels and CCD chip are in line with the optical axis of the telescope with allowable margins of shift and tilt estimated in the optical design software. Since the gravity and thermal effects will disturb the positions above, a proper design evaluation is necessary to achieve the tolerance limit estimated in optical designs.
	
\subsection{Finite Element Analysis (FEA) and design optimizations}

\begin{figure*}[!t]
\centering
\includegraphics[scale=0.4]{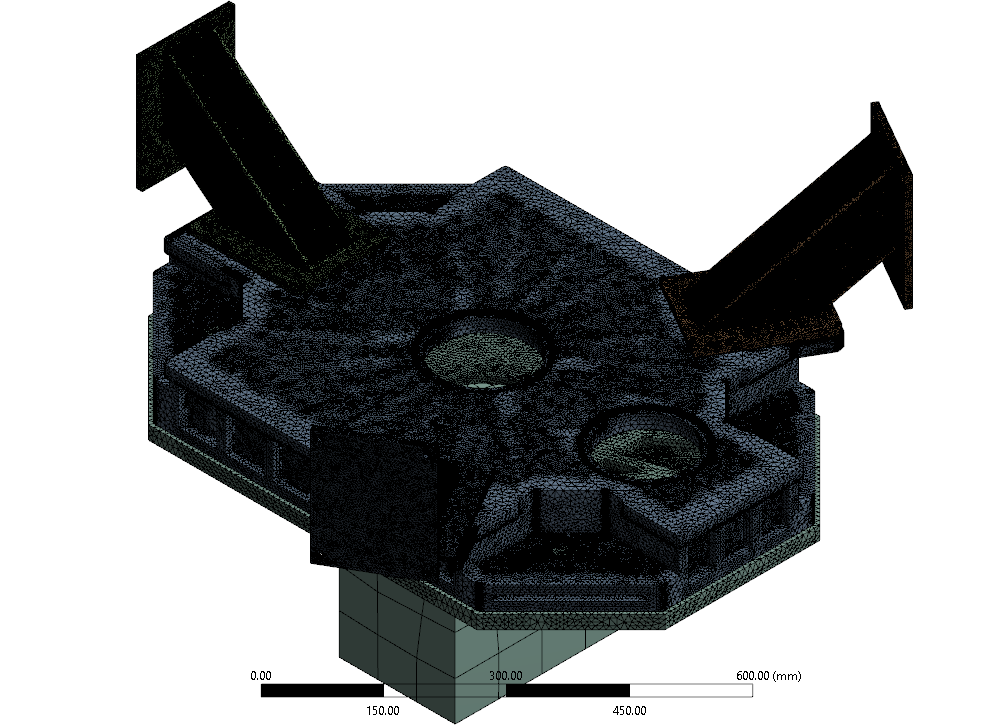}
\caption{System level finite element model with CCD, arm and filter wheels.}\label{simulated1}
\end{figure*}

The performance of the Imager instrument setup is strongly dependent on the suitable design of the housing structure, arms, positioning of filter wheels, on providing stable support to the CCD, on how well these designs take the effect of gravity, and also on thermal distortions. These effects are considered individually and through system-level analysis in FEA. The system model reflects the preliminary design incorporating an optimized ribbed housing structure topology and filter wheel weight, and camera dummy. The comprehensive iterations and systematic studies were carried out to select the position of arms and the number of arms required, also the integration and assembly point of view of these arms were finalized. Therefore, the three arms are designed in steel to avoid distortion and provide stiff support to the housing structure after FEA analysis (see Figure~\ref{simulated1}).

The housing structure was changed to aluminium casting to reduce the cost of manufacturing and the thickness of the body was reduced wherever possible. Further, ribs were provided to increase the strength of the housing structure. Brass inserts were incorporated to avoid wear and tear on threaded holes. Since the housing structure design optimization was performed on a finite element model, it was necessary to re-tune the housing structure to account for the gravity and thermal distortion effects that cause the early system to behave slightly differently in telescope elevation angles. The effect of CCD weight was incorporated by a coefficient of the frictional joint to the housing structure as an option available in Ansys software. The thickness of the housing structure, arm and filter wheel weight, etc. were also determined to keep the distortion within the optical design tolerance limits. The Filter position and CCD chip shift were determined within a pixel size.

\begin{figure*}[!t]
\centering
\includegraphics[scale=0.38]{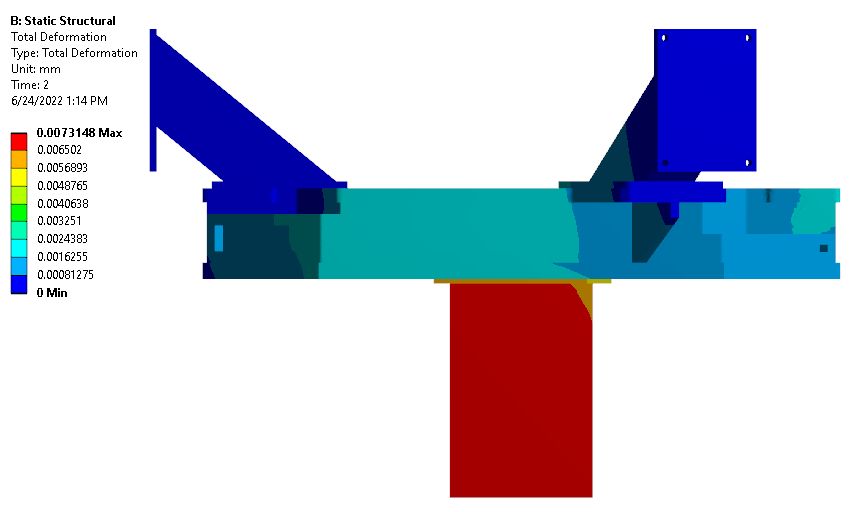}
\includegraphics[scale=0.42]{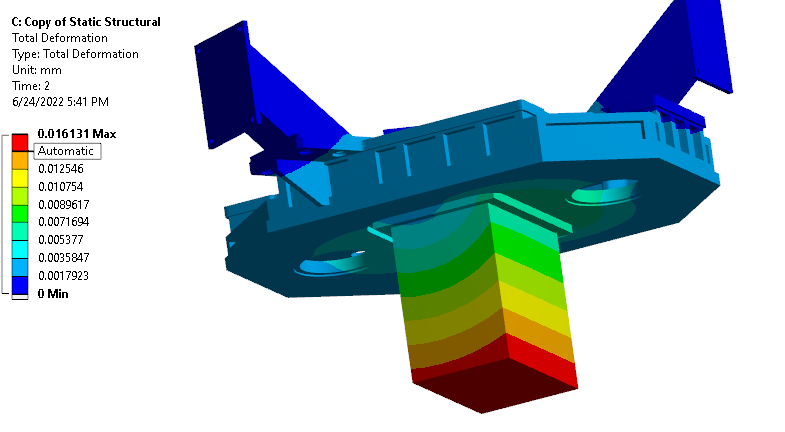}
\caption{Finite element analysis (FEA) estimates deflection of the Imager instrument during telescope pointing at zenith with zero elevation angle. The change of tilt values at chip /filter is within 10 arc minutes.}\label{simulated2}
\end{figure*}

The tilt values of the filters and CCD chip positions were determined within the optical tolerance of 10 arc minutes estimated in the optical design of \citet{Pandey2018a}. The stress and strain in designs were below the safe limit of material properties. Figure~\ref{simulated2} shows the FEA estimate deflection of the Imager instrument during the telescope pointing towards the zenith and zero elevation angle.

\section{Upgradations of the motorized filter wheel}
\label{filters}

\noindent 3.6m DOT Imager instrument consists of two filter wheels which are used for imaging at different wavelengths. First, filter wheel has six filter positions, namely $U$, $B$, $V$, $R$, $I$ and $C$ ($clear$) and similarly the second one also has six filter positions, namely $u$, $g$, $r$, $i$, $z$ and $c$ ($clear$) position. Motorized filter wheel controller has been developed and implemented for controlling the above-mentioned filter wheels system. The controller is based on a microchip PIC 18F4431 microcontroller and wheels are rotated with the help of stepper motors. Controller communicates with PC GUI on Ethernet to RS232 converter. 

\subsection{Filter wheel mechanism}

The circular filter wheel has gear teeth machined on the periphery, and it is coupled with a stepper motor through flexible coupling and gearbox. A combination of one Hall Effect vane sensor and one ferrous vane interrupter (aligned with each other) are mounted on each wheel at a different position for initial referencing. Sensors are mounted on the fixed part of the wheel and interrupters are mounted on the moving part of the wheel. Optical filters are housed in the wheel mechanism and the required positioning is obtained by rotating the wheel with the help of a stepper motor.

\subsubsection{Controller Architecture}

The controller is implemented using a microchip PIC 18F4431 microcontroller having Serial Communications Interface (SCI) and Universal Synchronous Asynchronous Receiver Transmitter (USART) module. Microcontroller has an ample number of digital I/Os and supports RS232 full duplex communication with the help of a TTL to RS232 level shifter. The firmware of the microcontroller has been developed in C language and the cross compiler has been used to convert the C program into compatible hex code.  RD4 and RD5 pins of PORTD are used for the wheel 1 \& 2 home position sensor. RB2, RB3, RB4, and RB5 pins of PORTB are used for controlling  the  stepper  motors.  RC6 (TX) and  RC7 (RX) pins  of  PORTC  are  used  for interfacing  with  the RS232  port  of  a  PC  through  the MAX232 transceiver.  RD6  \&  RD7 pins  of PORTD are used for the request of CW and ACW filter wheel - 2 rotation in manual mode, and R C0, R C1 and R C2 pins of PORT C are used for the request of both wheel homing, CW and ACW filter wheel - 1 rotation in  manual  mode.  Manual  operations  are performed  by  the  push  button switches mounted on the controller. 

\begin{figure}
\centering
\includegraphics[scale=0.364]{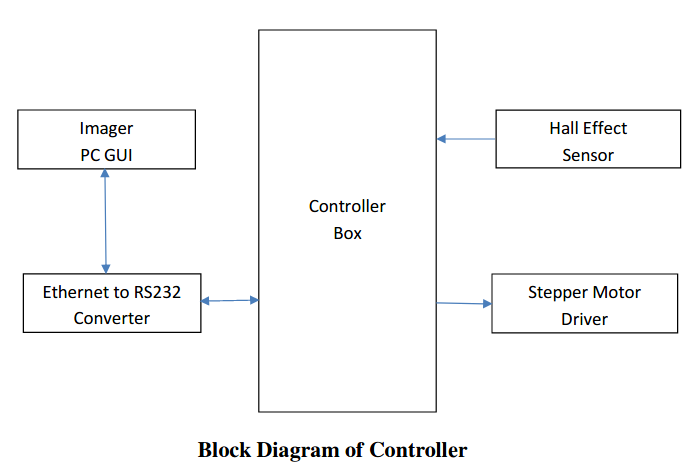}
\caption{Block diagram of motorized filter wheel controller.}
\label{FIGELEVEN}
\end{figure}

\begin{figure}[ht]
\begin{center}
\includegraphics[scale=0.5]{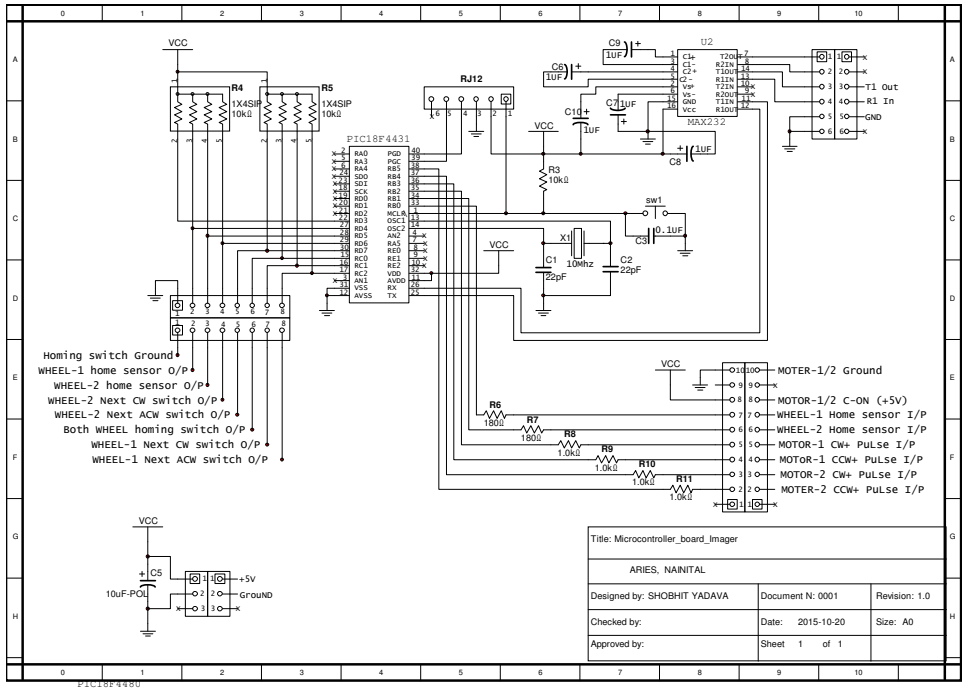}
\end{center}
\caption{Schematic diagram of the motorized filter wheel controller presently in use with the 4K$\times$4K CCD Imager.}
\label{FIGTWELVE}
\end{figure}

\begin{figure*}[!t]
\centering
\includegraphics[scale=0.5]{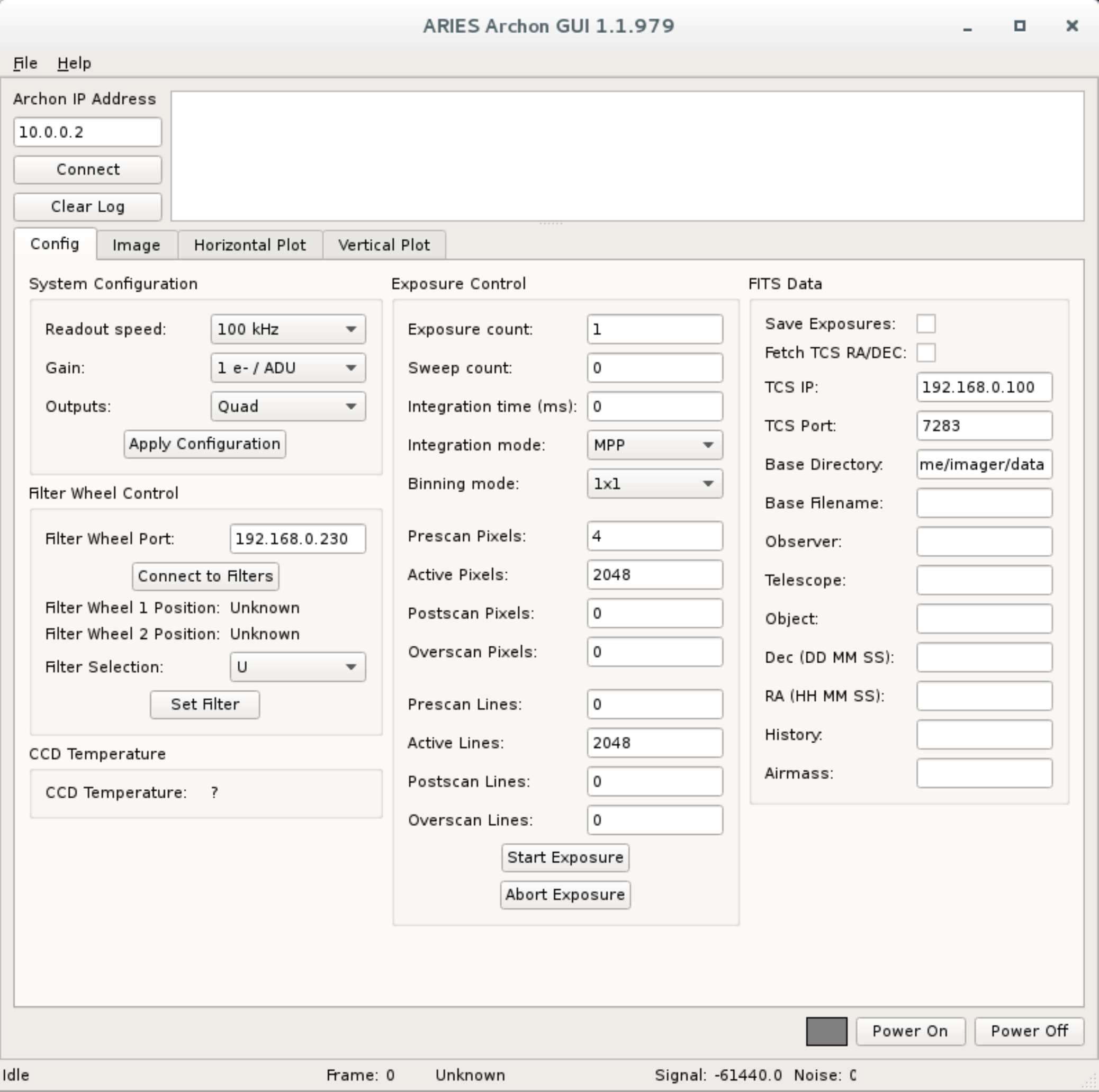}
\caption{A snapshot of the GUI utilized for the night observations using 4K$\times$4K CCD Imager.}\label{GUI}
\end{figure*}

The entire electronics system consists of a single control box which consists of a microcontroller and interfacing circuit along with a power supply. It also contains an Ethernet to RS232 converter and stepper motor driver units. The controller board has three ports, one for serial programming, a second for the PC interface and a third for connecting to sensors and the stepper motor driver unit of the filter wheel. The block diagram, schematic diagram of the Motorized Filter Wheel Controller, and the PC GUI are shown in Figure~\ref{FIGELEVEN}, Figure~\ref{FIGTWELVE} and Figure~\ref{GUI}, respectively. 

\subsection{Operations}
Once the system is set up and powered the wheel is continuously rotated (one after the other) till the home position is sensed. After homing controller keeps track of the number of steps and directions to know the exact position of the filter. The algorithm is written to follow the shortest path in either direction to reach the desired filter position. Wheels get locked at each filter position and therefore the motor needs not to be powered when the wheel is stationary. This ensures minimum heat dissipation from the motor near the focal plane. The angle between one filter to the other is 60 degrees at the wheel shaft. The gear ratio between the motor and to wheel shaft is 1:540. The stepper motor completes one rotation in 1000 steps.  It takes 90000 steps for moving from one filter to other.

Initially at the POWER ON both the wheels are positioned at their $clear$ positions. Once the PC GUI of the filter wheel is enabled, it sends a POWER ON status request to the controller and gets acknowledged. After POWER ON is acknowledged, it sends a current POSITION status request and gets the current position acknowledgement from the controller. Current POSITION status may be requested from the controller at any time once the power is energized. PC GUI has a provision to request all filter wheel position’s individual movement. If 1$^{st}$ filter wheel moves to any requested position then 2$^{nd}$ one will be moved to its $clear$ position. Any request received from a PC GUI will be acknowledged to the PC through Ethernet to RS-232 converter after the execution of the request by the controller. The controller will be ready for the execution of the next request after the execution of the previous request. During CCD readout time, the filter wheel's movement may be requested for the next required filter position, So that the filter wheels may reach its required position during the CCD readout time itself or it may save time to reach its required filter position. 

\subsection{Proposed modifications}
The present system is having only a combination of one sensor and one interrupter (aligned with each other) which is used for initial referencing. Filter wheel positioning is obtained through the counting of stepper motor pulses. The flexible coupling is used to mechanically couple the filter wheel and stepper motor with each other. Sometimes positioning error is observed due to loose/slippage coupling issues or count missing of stepper motor pulses and does not get displayed on the PC GUI. PC GUI only displays that the filter is in the requested position.

The modified system will be having two Hall Effect vane sensors and seven ferrous vane interrupters mounted on each wheel at a different position. Sensors will be mounted on a fixed part of the wheel and interrupters will be mounted on the moving part of the wheel. A combination of one sensor and one interrupter (aligned with each other) will be used for initial referencing. A combination of the remaining sensor and six interrupters (aligned with each other) will be used for exact position sensing/ pulse counting.

Firmware of the microcontroller is being developed which uses both stepper motor pulses and pulse from the exact position sensor for requested filter positioning and it will be implemented on a modified system. Errors due to count missing of stepper motor pulses will get corrected and errors due to loose/ slippage coupling issues will get displayed on the PC GUI if the filter is not reached on the requested position within the certain time limit. Once the error gets displayed on the PC GUI, coupling issues may be rectified accordingly.

\section{Fringe Correction}
\label{Fringr}

Back-illuminated thin CCD images are often affected by fringe patterns that are the results of the interference of monochromatic light within the CCD (Bernstein et al. 2017). Narrowband filters and broadband filters containing strong sky emission lines are typically affected by the fringes. Lines due to atmospheric O{\sc ii} and OH affect the red-end bandpasses of the CCD (wavelength $>$ 700 nm), i.e., mainly i- and z-bands (e.g., Gullixson 1992, Howell 2006, 2012). Though fringes add only a small additional flux to the image, their removal is a necessity for scientific and cosmetic reasons. 

Fringe patterns are also observed in the red-end bandpass images taken with the Imager. Removal of the fringes requires an extra step, where a description of the fringe pattern is scaled and subtracted from each image. Fringe patterns in the images are determined by the thickness variations of the CCD. Therefore, fringe patterns on the CCD are globally stable with time. It indicates that a single high signal-to-noise (S/N) fringe map may specify the fringe pattern. The fringe patterns can be generated through a series of night sky frames taken with a significant jitter pattern. To generate the fringe pattern, we median combine the night sky frames taken with a jitter pattern, so that only the fringe pattern is left. Once we have constructed a fringe map, it should be scaled to the intensity of the fringe pattern in each science image, and then subtracted. 

\begin{figure}
\centering
\includegraphics[scale=0.364, trim = 0 0 0 0, clip]{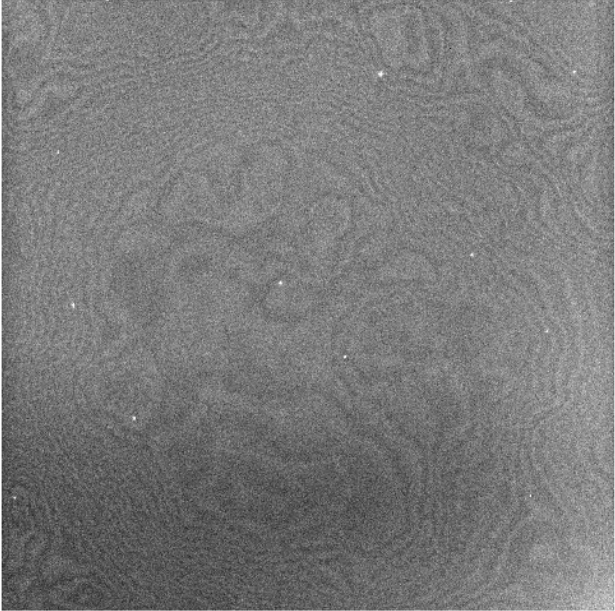}
\includegraphics[scale=0.33, trim = 0 0 0 0, clip]{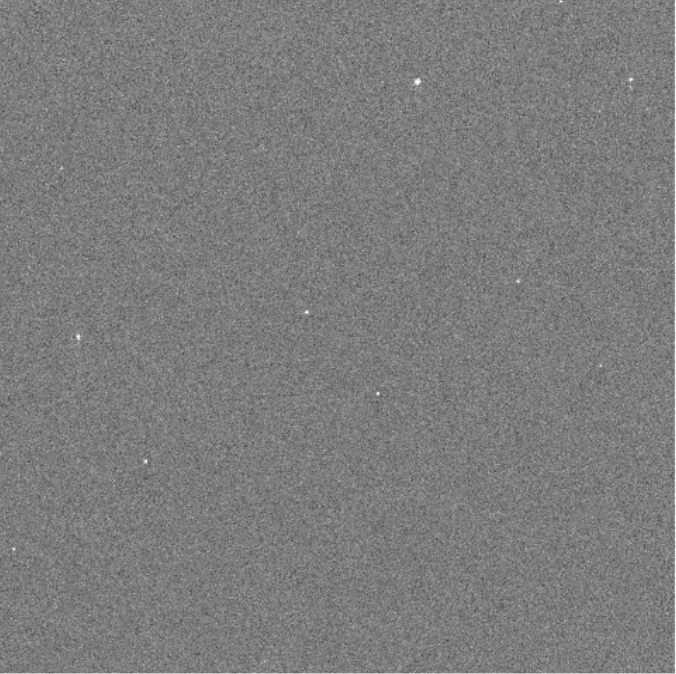}
\caption{An I-band image taken with the Imager before (left) and after fringe correction (right).}
\label{FIGTHIRTEEN}
\end{figure}
 
In general, the intensity of the fringe pattern becomes more prominent with increasing exposure time. For very short exposures these are hardly visible but for longer exposures, these clearly stand out of the background. Fringe removal becomes critically important, especially when dealing with the photometry of multiple or extended objects, or faint sources to provide properly uniform photometry across the image. So, first, the fringe pattern can be scaled by the exposure time. $IRAF$ tasks can be used for fringe correction. The $ccdproc$ task can also be used for the fringe correction. $IRAF$ package has the $mkfringecor$ task, which is used to combine frames to construct a fringe pattern. Another approach in $IRAF$ is first to scale the fringe map to globally minimize the difference between the map and object frames, and then use the $rmfringe$ and $irmfringe$ tasks in the $mscred$ package (Valdes, 1998). 
 
In Figure \ref{FIGTHIRTEEN}(left), we have shown an I-band raw image of a field. The fringes are clearly visible in the raw image and we found that the level of fringing in the image amounts to about 4-5\% modulation. To generate the fringe pattern, we took the blank sky frames in I-band. We used the $IRAF$ tasks $objmasks$, $sflatcombine$, and $rmfringe$ to generate the fringe pattern and apply the fringe correction to the raw image. After the fringe correction, the fringe pattern is completely removed (see Figure \ref{FIGTHIRTEEN} right panel) and the image is flattened.

\section{New Science Objectives}
\label{science}

The 4K$\times$4K CCD Imager is one of the most suitable instruments for the deep imaging of the galactic and extra-galactic astronomical sources in a set of ten broad-band filters (Bessel $UBVRI$ and SDSS $ugriz$). In addition, the longitudinal advantage of India and, particularly, the 3.6m DOT site is crucial in observing transients up to deeper limits \citep{Pandey2018a}. This section highlights some of the recent initiatives suited for imaging with the 4-m class optical telescopes and observed using the 4K$\times$4K CCD Imager.

\subsection{Extra-galactic Wolf-Rayet Stars}

Stars with the zero-age main sequence mass (M$_{ZAMS}$) of $\gtrsim$ 25 M$_\odot$ are supposed to evolve into WR stars \citep{Hamann2006}. The spectra of Wolf-Rayet (WR) stars exhibit features of heavy metals as their high temperature and strong stellar winds strip off the outer hydrogen envelopes. The outer envelope stripping could be through steady line-driven winds \citep{Puls2008}, eruptive episodes of mass loss \citep{Smith2006} and/or mass-transfer to a binary companion through Roche-Lobe overflow \citep{Yoon2010}. WR stars are thought to be the possible progenitors of most of the stripped-envelope supernovae (SESNe) and long gamma-ray bursts (GRBs); hence are very important in understanding these explosions. To study the WR stars and their environment, M101 (Messier 101 or NGC 5457 or Pinwheel Galaxy) situated in the Ursa Major constellation is one of the ideal spiral galaxies because of its face-on orientation and distance of $\approx$6.4 Mpc \citep{Shappee2011, Pledger2018} having $R$-band apparent magnitude of $\sim$7.76 mag as observed multiple times under the SDSS program. M101 also hosted a type Ia SN named SN 2011fe, which possibly originated from the death of a luminous red giant star \citep{Li2011}. Due to the large angular size and proximity of M101, stellar content and their evolution in the spiral arms could be studied better.

\begin{figure*}[!t]
\centering
\includegraphics[scale=0.32]{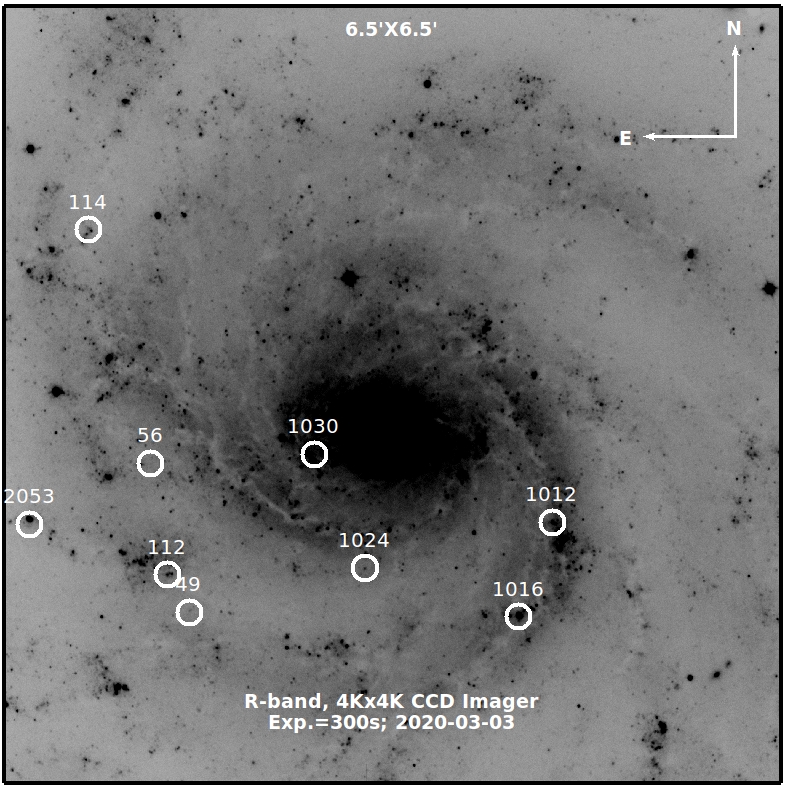}
\caption{$R$-band frame of M101 observed on 2020-03-03 with a field-of-view. of $\sim$6.5$'$ $\times$6.5$'$ and exposure time of 300 sec with the 4K$\times$4K CCD Imager. Out of 15 regions hosting WR stars (as reported in \citet{Pledger2018}), 9 regions were identified in our observations demonstrating capabilities of the 3.6m DOT for such studies.}\label{FIGFORTEEN}
\end{figure*}

We also attempted to detect the WR stars situated in the crowded spiral arms of M101 with the help of the 4K$\times$4K CCD Imager. The $R$-band image of M101 with a field of view of $\sim$6.5$' \times$6.5$'$, having an exposure time of 300 sec and observed on 2020-03-03 is shown in Figure~\ref{FIGFORTEEN}. Out of 15 WR stars in M101 detected by \citet{Pledger2018}, we are able to locate 9 WR stars (IDs: 49, 56, 112, 114, 1012, 1016, 1024, 1030 and 2053) that are highlighted with circles. Four of the nine encircled WR stars are well resolved and detected above the 3-sigma limit (IDs: 49, 112, 114, 1012, and 1024; see Figure~\ref{FIGFORTEEN}), hence ideal for performing the photometry. Larger apertures are needed to estimate the total flux of the stars. But placing a larger aperture can also lead to more flux from the sky and also from the extended galaxy flux, which can lead to more considerable uncertainties in the fluxes of stars. Therefore, initially, we chose small apertures (1 $\times$ FWHM) and applied the aperture correction later. We chose multiple isolated and bright field stars to estimate the aperture correction. First, the fluxes of the field stars were measured for 1 $\times$ FWHM and 4 $\times$ FWHM aperture radii and calculated the aperture correction factor. As the WR stars mentioned above are in crowded galaxy arms, the PSF photometry is performed based on the PSF radius of 1 $\times$ FWHM and then applied the aperture correction term to estimate the total fluxes. The above-discussed tasks are performed using the Python scripts hosted on RedPipe \citep{Singh2021}. The same standard stars used for the aperture correction are also used for the calibration based on their $R$-band magnitudes quoted in the USNO-B1. The final estimated magnitudes of the WR stars are tabulated in Table~\ref{tab:mags_comparison}, shown in bold. The $R$-band magnitudes for the above discussed WR stars are reported for the first time and not available in the literature hence compared with the $F555W$ band magnitudes estimated by \citet{Pledger2018}. For the above-discussed four WR stars, the colour terms for $F555W$ and $R$-band magnitudes are nearly the same with a scatter of $\sim$0.2 mag, which may be attributed to possible temporal variability with time. The ID numbers of each WR star mentioned in this analysis are similar to those discussed in Tables 3, 4, and 5 of \citet{Pledger2018} and more details about these stars are published therein. The present analysis demonstrates the capabilities of the 4K$\times$4K CCD Imager plus 3.6m DOT to resolve the extra-galactic stars embedded in their parent galaxies.

\begin{table}
  \caption{The $R$-band magnitudes of the four WR stars with IDs 49, 112, 114, and 1012 are shown in bold. The estimated magnitudes are also compared with the $F555W$-band (central wavelength = 5308.4 $\AA$ and width = 1565.4 $\AA$) magnitudes reported by \citet{Pledger2018}.}
\centering
\begin{tabular}{l@{\hspace{12mm}}c@{\hspace{12mm}}c@{\hspace{12mm}}c@{\hspace{12mm}}c@{\hspace{12mm}}}
\hline
ID & m$_{F555W}$ & err & R-band & err  \\
\hline
49  &  24.33 & 0.02  & 21.79 & 0.16 \\
112 &  22.71 & 0.04  & 20.23 & 0.14 \\
114 &  20.60 & 0.04  & 18.25 & 0.11 \\
1024 & 23.81 & 0.03  & 21.50 & 0.15 \\
\hline

\end{tabular}
\label{tab:mags_comparison}

\end{table}

\begin{table}
  \caption{The R-band magnitudes of the four galaxies (IDs 1, 2, 3, and 4) are tabulated (in bold) and are also compared with the SDSS $r$-band magnitudes.}
\centering
\begin{tabular}{l@{\hspace{12mm}}c@{\hspace{12mm}}c@{\hspace{12mm}}c@{\hspace{12mm}}c@{\hspace{12mm}}}
\hline
ID & SDSS $r$-band & err & Bessel R-band & err  \\
\hline
1 &  18.45 & 0.02  & 18.13 & 0.08 \\
2 &  18.30 & 0.01  & 17.90 & 0.06 \\
3 &  20.68 & 0.06  & 20.18 & 0.14  \\
4 &  20.87 & 0.06  & 20.21 & 0.15 \\
\hline

\end{tabular}
\label{tab:galaxy_mags_comparison}

\end{table}

\begin{figure*}[!t]
\centering
\includegraphics[scale=0.35]{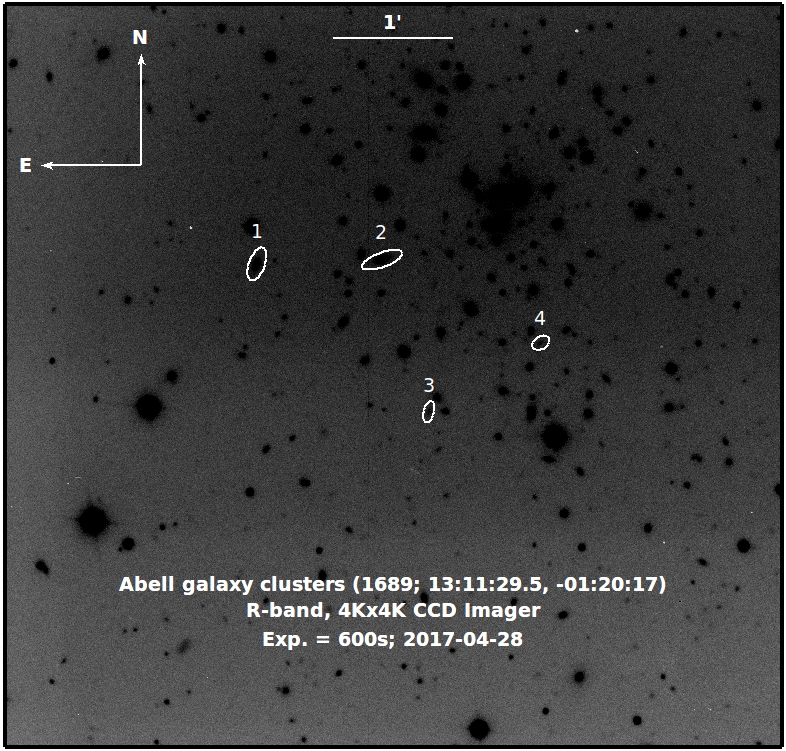}
\caption{Stacked $R$-band frame (3$\times$200 sec = 600 sec) of the ACO 1689 observed using the 4K$\times$4K CCD Imager on 2017-04-28.}\label{FIGSIXTEEN}
\end{figure*}

\subsection{Abell Galaxy Cluster}\label{sec:abell}

Galaxy clusters are gravitationally bound structures containing hundreds of galaxies and X-ray-emitting hot intracluster plasma, but their mass budget is dominated by dark matter.  One of the well-known survey catalogues, Abell, includes nearly 4000 galaxy clusters up to $z$ $\sim$0.2 \citep{Abell1958, Abell1989}. The galaxy cluster with the Abell, Corwin and Olowin (ACO) number 1689 (RA: 13h:11m:29.5s, DEC: -01$^\circ$20$'$17$''$; J2000) and situated in the $Virgo$ constellation is exciting to investigate because of some of its peculiar characteristics. ACO 1689 is one of the most massive and giant galaxy clusters (ACO 1689 contains around 0.16 million globular clusters, the largest population even found in any galaxy cluster) observed so far \citep{Tyson1995}, which presents significant gravitational lensing \citep{Limousin2007, Ghosh2022}. Lensing effects in galaxy clusters are more powerful in central regions $<$ 0.3 Mpc having high concentrations of both baryonic and dark matter often producing multiple images and serving as suitable natural laboratories to study possible relations between dark matter and baryonic ones and underlying physical processes within these regions. Additionally, strong gravitational lensing within galaxy clusters are also powerful probes to study high red-shift universe as clusters can magnify faint distant background sources, making them observable with the help of moderate to large size telescopes
\citep{2010CQGra..27w3001B}. ACO 1689 is one of the well-studied clusters to be revealed with more than $>$ 100
multiple images from 30 sources \citep{2005ApJ...621...53B}.

In the current analysis, we present the $R$-band image of the ACO 1689, observed on 2017-04-28 with the help of the 4K$\times$4K CCD Imager at the 3.6m DOT as a part of calibration test observations (see Figure~\ref{FIGSIXTEEN}). After the pre-processing and alignment, three individual $R$-band images, each of 200 sec, were stacked to achieve a better signal-to-noise ratio. To highlight the capability of 3.6m DOT + 4K$\times$4K CCD Imager in observing the extended objects, we also performed the photometry of four random member galaxies with a range of brightness (up to $\sim$ 20.2 mags), marked with IDs 1, 2, 3 and 4 in Figure~\ref{FIGSIXTEEN}. The photometric analysis is accomplished using the {\tt SEP} (Source-Extractor), which is a python package suitable for performing faint-galaxy photometry because of its ability to measure accurate PSF and galaxy model fitting \citep{Bertin1996, Barbary2016}. The estimated $R$-band magnitudes for the above-discussed four galaxies are tabulated in Table~\ref{tab:galaxy_mags_comparison} and also compared with those of SDSS $r$-band magnitudes. The $R-r$ colour terms for the four galaxies discussed here are consistent with the colour equation discussed by \citet{Jordi2006}. Our observations based on deep imaging of the ACO 1689 in the $R$-band demonstrate the capabilities of the 3.6m DOT for such studies. We have been able to detect many galaxies and lensed sources within the central region of the cluster down to a limiting magnitude of $\sim$21 mag in the $R$-band in an exposure time of 600s. The detailed analysis of these observations and the photometric analysis procedures are ongoing and will soon be published elsewhere.

\subsection{Gamma-Ray Bursts and host galaxies}

GRBs are among the most exotic and luminous ($\rm L_{\gamma, iso}$ $\sim$ $10^{48}$ $-$ $10 ^{54}$ erg/s) phenomena studied in modern astronomy \citep{2004RvMP...76.1143P}. They have two distinct emission phases; one is the prompt emission (the initial burst phase, peak at sub-MeV energy range), followed by a long-lived multiwavelength afterglow phase \citep{2015PhR...561....1K}. The longitudinal advantage of 3.6m DOT and deep imaging capabilities of 4K$\times$4K CCD Imager together provide a unique opportunity to perform deep and longer follow-up observations of the optical counterparts/ associated host galaxies of GRBs. Recently, 3.6m DOT+ 4K$\times$4K CCD Imager discovered a few interesting results such as the detection of orphan and dark afterglows \citep{2021arXiv211111795G}, the detection of most delayed optical flare (originated due to refreshed shock) observed from any GRB so far \citep{2022MNRAS.513.2777K}, detection of long GRB (GRB 211211A) from the binary merger \citep{2021GCN.31299....1G}. In addition, we also studied a sample of host galaxies of GRBs observed using the 3.6m DOT+ 4K$\times$4K CCD Imager and compared the results with a larger sample of normal star-forming galaxies. We noted that the physical properties such as star-formation rate, mass, and specific star-formation rate of the host galaxies of GRBs are more similar to the normal star-forming galaxies at the higher redshifts \citep{2022arXiv220513940G}. 

\begin{figure*}[!t]
\centering
\includegraphics[scale=0.4]{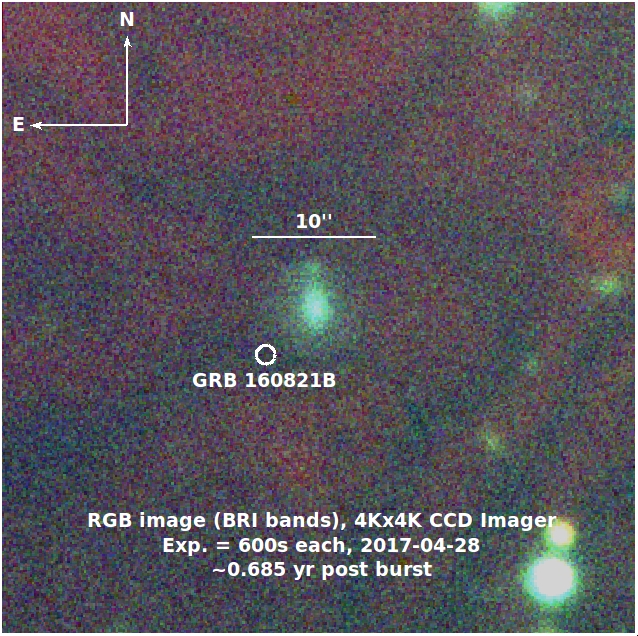}
\caption{An $RGB$ image prepared using the $I$, $R$, and $B$ band frames of the  GRB 160821B field observed on 2017-04-28 using the CCD imager. Each frame of $BRI$ bands exhibited exposure times of 600 sec. The position of GRB 160821B is also marked with a circle. GRB~160821B was not detected up to 3$\sigma$ upper limit of B $\approx$22.2 mag at the epoch of our observations.}\label{FIGFIFTEEN}
\end{figure*}

In this work, we report the host galaxy observations of one of the interesting and nearest ($z$ = 0.162) short GRB (GRB 160821B). GRB 160821B was triggered and localized by the {\it Swift} mission  \citep{2016GCN.19833....1S}. \citet{2019MNRAS.489.2104T, 2019ApJ...883...48L} performed the analyses of the multiwavelength afterglow data and revealed an optical-infrared kilonova emission, powered by heavy-element nucleosynthesis in a binary neutron star (BNS) merger. Later on, \citet{2021ApJ...908...90A} reported the detection of very high energy emission (VHE) from GRB 160821B at a significance of $\sim$ 3 $\sigma$ using MAGIC telescope, making it the first short GRB with TeV emission. The detection of VHE emission from short bursts is useful to constrain their possible progenitors and enhance our understanding towards BNS \citep{2021ApJ...908...90A}. 

We observed the field of GRB 160821B using 3.6m DOT+ 4K$\times$4K CCD Imager in multiple broadband filters ($BRI$) $\sim$0.685 years post burst. Figure \ref{FIGFIFTEEN} shows the finding chart (RGB image) of the field of GRB 160821B took using 3.6m DOT + 4K$\times$4K CCD Imager. We noted that the associated afterglow/kilonova emission at the position of GRB 160821B has significantly faded, consistent with the typical decay rate. However, we clearly detected the associated host galaxy of GRB 160821B in all the filters. We also constrain the $B$, $R$ and $I$-band magnitudes of the host galaxy of GRB 160821B utilising the {\tt SExtractor} as discussed in Section~\ref{sec:abell}. The GRB host's estimated $B$, $R$ and $I$-band magnitudes ($\sim$19.5 $\pm$ 0.10, 19.16 $\pm$ 0.06 and 19.02 $\pm$ 0.09 mags, respectively) are consistent with those published earlier ($B$ $\sim$19.6; \citealt{2019MNRAS.489.2104T} and $R$ $\sim$19.2; \citealt{2016GCN.19834....1X}).

Such deep photometric observations of the host galaxy are crucial to estimate the metallicity, mass, ages of the galaxies, star-formation rates, and other physical characteristics of environments of GRBs and hence progenitors. Recently, \citet{2022arXiv220601763F, 2022arXiv220601764N} studied the host galaxies properties of a larger sample of short GRBs including the host of GRB 160821B and calculated the following stellar population parameters for GRB 160821B using spectral energy distribution modelling of the host: log(M/M$_{\odot}$) = 9.24$^{+0.01}_{-0.01}$, log(Z/Z$_{\odot}$) = 0.1$^{+0.05}_{-0.05}$,  $t_{gal}$ = 0.58$^{+0.02}_{-0.02}$ Gyr, $A_{V}$ = 0.01$^{+0.01}_{-0.01}$ mag, and SFR = 0.24$^{+0.01}_{-0.01}$ M$_{\odot}$/yr, where log(M/M$_{\odot}$), log(Z/Z$_{\odot}$), $t_{gal}$, $A_{V}$, and SFR are stellar mass formed, stellar metallicity, age of the galaxy, rest-frame dust attenuation in the host, and star-formation rate, respectively. The calculated physical parameters for GRB 160821B are in agreement with typical host galaxies of short GRBs. The afterglow position of GRB 160821B is located at $\sim$ 16 kpcs projected physical distance from the centre of the associated spiral host galaxy, such large offset values are observed in the case of short GRBs only \citep{2019MNRAS.489.2104T}. In the near future, we plan to continue deep and long follow-up photometric observations of optical afterglows/ host galaxies of new exciting transients during LIGO O4 run and later on visible in the DOT sky utilizing the unique capabilities of 3.6m DOT+ 4K$\times$4K CCD Imager and other back-end instruments.

\section{Summary}
\label{summary}
This article summarises details of various sub-components of the 4K$\times$4K CCD Imager not published so far. The present details were not covered in the earlier published papers related to the instrument \citep{Pandey2018a, Kumar2022a}.  Issues noticed in the images, like partial vignetting towards the edges, were understood and resolved with the help of a modified light baffle now installed with the imager and discussed in detail in section 2. Section 3 summarised the mechanical design and related analysis like FEA and design optimizations of the filter housing and other sub-components in-house in the ARIES workshop. To maximize the repeatability of the filter wheel movement and to minimize possible effects like partial vignetting due to associated errors in filter wheel movement, the proposed design modifications are discussed in section 4, which will be implemented with the new backup electronics soon. In section 5, we present our efforts to minimize the fringe patterns usually observed in near-IR filters like $I,i,z$ and have attempted to quantify the possible effects on the measurements of the brightness of celestial objects observed in such filters, a useful exercise for observers using near-IR filters for various objects with the 3.6m DOT. These results are indeed useful for calibration and reference for any back-end instrument to be developed in the near future for the 3.6m DOT. In section 6, we briefly discuss some of the new science initiatives suited for the 3.6m DOT, considering the longitudinal advantage of the place. We also presented the multiband photometric analysis of some faint point objects and faint galaxies, indicating the capabilities of the 4K$\times$4K CCD Imager as a direct imaging instrument available at the axial port as and when required. The science cases discussed here further demonstrate the importance of this national facility and are only a few among many more published elsewhere, adding value towards the possible range of front-line scientific needs of the community.

\section*{Acknowledgements}
SBP, AA, and RG acknowledge BRICS grant {DST/IMRCD/BRICS/PilotCall1/ProFCheap/2017(G)} for the financial support. RG and SBP acknowledge the financial support of ISRO under AstroSat archival Data utilization program (DS$\_$2B-13013(2)/1/2021-Sec.2). AA acknowledges funds and assistance provided by the Council of Scientific \& Industrial Research (CSIR), India with file no. 09/948(0003)/2020-EMR-I. This research is based on observations obtained at the 3.6m Devasthal Optical Telescope (DOT) which is a National Facility run and managed by Aryabhatta Research Institute of Observational Sciences (ARIES), an autonomous Institute under the Department of Science and Technology, Government of India. SBP, AK and the imager team thankfully acknowledge support and assistance from the DOT team and the whole technical staff of ARIES to make the facility working and usable for the community.

\bibliographystyle{apj}
\bibliography{wsjai}

\end{document}